\documentclass[twocolumn,showpacs,preprintnumbers,amsmath,amssymb,pra]{revtex4}
\usepackage{graphicx}
\usepackage{dcolumn}
\usepackage[tight]{subfigure}
\usepackage{amsmath}
\usepackage{verbatim}
\usepackage[usenames,dvipsnames]{color}
\usepackage{bm} 
\usepackage{bbm}
\newcounter{bibmaxnames}
\newcommand{\op}[1]{#1}
\newcommand{\tmem}[1]{{\em #1\/}}
\newcommand{\tmop}[1]{\ensuremath{\operatorname{#1}}}

\newcommand{\Cite}[1]{~\onlinecite{#1}}
\definecolor{grey}{rgb}{0.75,0.75,0.75}
\definecolor{orange}{rgb}{1.0,0.5,0.5}
\definecolor{brown}{rgb}{0.5,0.25,0.0}
\definecolor{pink}{rgb}{1.0,0.5,0.5}
\newcommand{\tmfloatcontents}{}
\newlength{\tmfloatwidth}
\newcommand{\tmfloat}[5]{
  \renewcommand{\tmfloatcontents}{#4}
  \setlength{\tmfloatwidth}{\widthof{\tmfloatcontents}+1in}
  \ifthenelse{\equal{#2}{small}}
    {\ifthenelse{\lengthtest{\tmfloatwidth > \linewidth}}
      {\setlength{\tmfloatwidth}{\linewidth}}{}}
    {\setlength{\tmfloatwidth}{\linewidth}}
  \begin{minipage}[#1]{\tmfloatwidth}
    \begin{center}
      \tmfloatcontents
      \captionof{#3}{#5}
    \end{center}
  \end{minipage}}

\newcommand{\Figure}[2]{
  \begin{figure}[ht]
    \includegraphics[width=1.0\linewidth]{#1}
    \caption{#2}
  \end{figure}
}
\expandafter\newcommand\csname Figure2\endcsname[3]{
\begin{figure}[ht]
  \includegraphics[width=0.9\linewidth]{#1}
  \\
  \includegraphics[width=0.9\linewidth]{#2}
  \caption{#3}
\end{figure}
}

\newcommand{\new}[1]{\textcolor{blue}{#1}}
\newcommand{\newer}[1]{\textcolor{blue}{#1}}
\newcommand{\New}[1]{\textcolor{blue}{#1}}

\newcommand{\cut}[1]{\textcolor{magenta}{CUT: #1}}

\renewcommand{\cut}[1]{}
\renewcommand{\new}[1]{{#1}}
\renewcommand{\newer}[1]{{#1}}
\renewcommand{\New}[1]{{#1}}

\renewcommand{\vec}[1]{{\bf{#1}}}
\renewcommand{\op}[1]{{\hat{#1}}}

\begin{document}

\title{
  Spin resonance without spin splitting
}

\author{M. Hell$^{(1,2)}$}
\author{B. Sothmann$^{(3)}$}
\author{M. Leijnse$^{(4)}$}
\author{M. R. Wegewijs$^{(1,2,5)}$}
\author{J. K\"onig$^{(6)}$}
\affiliation{
  (1) Peter Gr{\"u}nberg Institut,
      Forschungszentrum J{\"u}lich, 52425 J{\"u}lich,  Germany
  \\
  (2) JARA- Fundamentals of Future Information Technology
  \\
  (3) D{\'e}partement de Physique Th{\'e}orique, Universit{\'e}
  de Gen{\`e}ve, CH-1211 Gen{\`e}ve 4, Switzerland
  \\
  (4)
  Solid State Physics and Nanometer Structure Consortium (nmC@LU),
  Lund University, 221 00 Lund, Sweden
  \\
  (5) Institute for Theory of Statistical Physics,
      RWTH Aachen, 52056 Aachen,  Germany
  \\
  (6) Theoretische Physik and CENIDE, Universit{\"a}t 
  Duisburg-Essen, 47048 Duisburg, Germany
}
\date{\today}

\begin{abstract}
We predict that a single-level quantum dot without discernible splitting of its spin
 states develops a
spin-precession resonance in charge transport when embedded into a spin
 valve.  \newer{The resonance occurs in the generic situation of Coulomb
 blockaded transport with ferromagnetic leads whose polarizations deviate from perfect antiparallel alignment.}
The resonance appears when \new{electrically tuning} the interaction-induced
 exchange field perpendicular to one of the \new{polarizations} -- 
a simple condition relying
 on vectors in contrast to usual resonance conditions associated with energy splittings.
The spin resonance can be detected by stationary $dI/dV$ spectroscopy and by oscillations in
the time-averaged current using a gate-pulsing scheme. 
The generic noncollinearity of the ferromagnets
and junction asymmetry allow for an all-electric determination of
 the spin-injection asymmetry, the anisotropy of
spin relaxation, and the magnitude of the exchange field.
 We also investigate the impact of a nearby superconductor on the
resonance position. Our simplistic model turns out to be generic for a
 broad class of coherent few-level quantum systems.
\end{abstract}
\maketitle
\section{Introduction}\label{sec:spr-intro}
Gaining fast, coherent control over a few spins or even a single spin is at the heart of current
experimental efforts in both spintronics \cite{Loth10,Besombes12,Khajetoorians13}
and solid-state quantum computing
\cite{Petta05,Koppens06,Nowack07,Bluhm10}. Single-molecule magnets in gateable nanojunctions \cite{Fernandez-Rossier07,Bogani08,Zyazin10,Urdampilleta11}
 or adatoms and molecules manipulated by STM \cite{Heinrich04, Loth10,Kahle12,Heinrich13} provide a bottom-up approach
to achieve this goal. Promising top-down routes combine conventional
spin valves \cite{Julliere75,Baibich88,Binasch89,Camley89,Slonczewski89}
with nanoscale quantum dot (QD)
devices \cite{Seneor07,Sergueev02,Wilczynski05,Fransson05,Weymann05,Pedersen05,Baumgaertel11}.
Such coherent quantum systems are typically manipulated through resonance techniques, e.g.,
by electromagnetic pulses \cite{Koppens06,Nowack07}.
In general, this requires that the frequency
of the applied pulses matches the splitting of, e.g., a two-level system. 
In this paper we predict
that quite generically resonances can appear in systems with quasi-degenerate levels
that do not involve such a matching to a splitting. \new{Instead,
a condition involving \emph{vectors} has to be satisfied.}\\
We illustrate this for a QD embedded in a noncollinear spin valve,
a specific example relevant for spintronics and spin-based quantum computation.
It leads to an unexpected, strongly gate-voltage dependent feature in
the stationary nonlinear conductance ($d I / d V_b$)
extending all across the Coulomb blockade regime.
It arises under nonequilibrium conditions but disappears upon
reversing the bias voltage.
Strikingly, it can appear at voltages much larger or smaller than any
of the naively expected energy scales, showing that it does not fit into the
usual classification of resonances.
All these features distinguish this resonance from known effects in the
Coulomb blockade regime \cite{Weymann05,Usaj01,Weymann05b}, including
those due to inelastic
cotunneling resolving
excitations \cite{Zumbuehl04,Paaske06,Kirsanskas12}, the Kondo
effect \cite{Pasupathy04kondo,Hamaya07,Hauptmann08,Gaass11}, and
\New{another zero-bias anomaly specific to QD spin valves \cite{Weymann05,Weymann05b}}.\\
The anomalous resonance we predict here relies on the coherent precession of a single spin
that is driven by the Coulomb interaction-induced \tmem{exchange field} \cite{Koenig03,Martinek03a,Koenig05rev,Choi04,Braun04set,Martinek05,Gaass11}.
The exchange field is a generic renormalization
effect \cite{Holm08,Splettstoesser12a,Sobczyk12,Kirsanskas12,Misiorny13} \newer{arising from quantum fluctuations of QD
electrons into the attached ferromagnets. This leads
to a spin-dependent level shift, i.e., an effective magnetic field, because the tunneling rates into
the ferromagnets are spin-dependent.}
While
this exchange field has been measured for
strong tunnel coupling $\Gamma$ as an \New{induced} level
splitting for collinear polarizations \cite{Pasupathy04kondo,Hamaya07,Hauptmann08,Gaass11},
the sharp resonance that we predict here appears for moderate tunnel couplings when this
splitting \emph{cannot} be resolved.
\newer{\New{In this case}, the exchange field can still have an impact under the additional requirement that the
rotational symmetry is broken completely by a noncollinear magnetic configuration
of the spin valve. Here, each ferromagnet
induces a contribution to the exchange field along its
polarization, which strongly depends on the applied voltages. This adds a \New{tunable} component to the exchange field that is
perpendicular to the injected spins. This induces a spin precession that}
results in measurable consequences for the stationary
conductance \cite{Koenig03,Braun04set,Braun05a} and
the noise spectrum \cite{Braun06noise,Sothmann10}, also for hybrid
setups with a superconductor \cite{Sothmann10c}.\\
\newer{However, \New{the features discussed so far} change on large voltage scales in
contrast to the sharp resonance presented in this work.}
This relates to the limitation of these prior works to the sequential
tunneling regime where the electron dwell times $1/\Gamma$ are too small
for single spins to precess by a large angle.
\New{To find a sharp resonance one needs} a suppression of the spin decoherence,
which is achieved in our case by an exponentially small leading-order $\Gamma$
contribution due to the Coulomb interaction $U$.
Our spin resonance thus appears in the Coulomb blockade regime of a
\new{QD spin valve}
where the spin decoherence is limited by higher-order contributions
$\propto \Gamma^2 / U$, while
the spin-precession period is still dominated by the leading-order $\Gamma$
exchange field \cite{Weymann07c}.\\
Only few studies
address spin-precession effects in the
Coulomb-blockade regime \cite{Rudzinski05,Weymann07c}. \New{What has
been overlooked in those works is} that a simple QD spin valve already has built-in
capabilities for single-spin operations through the gate-voltage control over the exchange field
\emph{direction} in the fixed, nearly antiparallel configuration.
We show that the resulting spin resonance can be exploited in a gate-pulsing scheme to provide single-spin control
for quantum-gate operations.
Time-averaged current measurements directly probe the underdamped spin precession.\\
The paper is organized as follows: In
Sec. \ref{sec:modelandkineq}, we first introduce
the QD spin-valve model under consideration and discuss our quantum
master equation approach to describe the dynamics of the QD system. Based on
the solution of these equations, we compute the stationary conductance that
exhibits the above-mentioned spin resonance. We substantiate the simple
resonance condition in Sec. \ref{sec:resonance} and identify relevant parameter combinations that
characterize the resonance features (position and width). We further suggest
procedures to extract these parameters from experimental data in order to characterize QD
spin valves. Next, we \New{propose} in Sec. \ref{sec:larmor} a simple
gate-pulsing scheme, which \New{is shown to} reveal the underdamped
spin precession occurring near the spin resonance. Finally, we summarize our
findings in Sec. \ref{sec:spr-summary} and
\New{argue} that the resonance mechanism described here is generic for a broad
class of coherently evolving quantum systems renormalized through their
environment.
\section{Model and kinetic equations}\label{sec:modelandkineq}
\New{The spin resonance appears in the simplest QD spin-valve model one can think of, which is introduced in Sec. \ref{sec:sprmodel}.
This is remarkable since this model of an interacting, single,
spin-degenerate orbital level, which is tunnel coupled to two
noncollinearly polarized ferromagnetic leads, has been studied quite intensively.
In Sec. \ref{sec:notation}, we discuss the quantum
master equation for the QD density operator $\rho$, which
is required to address this spintronic effect.} \newer{Based on the solution of these equations, we
derive the current that exhibits the spin-resonance feature.}
\subsection{Complete breaking of rotational symmetry in quantum-dot spin
valves}\label{sec:sprmodel}
The system under study, see Fig. \ref{fig:model}, consists of a QD,
which is tunnel coupled to two
ferromagnetic leads $r$, labeled with $r = s (\tmop{ource}), d (\tmop{rain})$.
The Hamiltonian reads:\\
\begin{eqnarray}
  H_{\tmop{tot}} & = & H + \sum_{r = s, d} H_r + H_T .  \label{eq:htot}
\end{eqnarray}
The QD is modeled by a single, spin-degenerate, interacting orbital level,
\begin{eqnarray}
  H & = & \sum_{\sigma} \varepsilon d^{\dag}_{\sigma} d_{\sigma} + U
  d^{\dag}_{\uparrow} d_{\uparrow} d^{\dag}_{\downarrow} d_{\downarrow}, 
  \label{eq:hd}
\end{eqnarray}
where $d_{\sigma}^{\dag}$ ($d_{\sigma}$) are fermionic field operators that
create (annihilate) electrons with spin $\sigma$ in the QD. The QD Hamiltonian
(\ref{eq:hd}) is spin {\tmem{isotropic}}, that is,
\begin{eqnarray}
  [ H, \op{S}_i ] & = & 0, 
\end{eqnarray}
where
\begin{eqnarray}
  \op{S}_i & = & \sum_{\sigma \sigma'} \tfrac{1}{2} (\sigma_i)_{\sigma
  \sigma'} d^{\dag}_{\sigma} d_{\sigma'}  \label{eq:sd}
\end{eqnarray}
is the $i$th Cartesian component ($i = x, y, z$) of the spin vector operator
and the $\sigma_i$ denote the Pauli matrices. The spin isotropy of the QD
model implies that the two spin states are degenerate.\\
By contrast, the spin symmetry is broken in the ferromagnets, held at
equal temperature $T$ and
different \new{electro}chemical potentials $\mu_{s (d)} = \pm V_b \text{/} 2$, with Hamiltonian
\begin{eqnarray}
  H_r & = & \sum_{k \sigma}^{} \varepsilon_{r k \sigma} c_{r k \sigma}^{\dag}
  c_{r k \sigma},  \label{eq:spr-hr}
\end{eqnarray}
where $c_{r k \sigma}^{\dag}$ ($c_{r k \sigma}$) are fermionic field operators
that create (annihilate) electrons in single-particle states
${ |} r k \sigma { \rangle} = { |} r k
{ \rangle} \otimes { |} \sigma
{ \rangle}_r = c^{\dag}_{r k \sigma} { |}
0 { \rangle}$ of lead $r$.
The spin-quantization axis is chosen for each ferromagnet \newer{along
its polarization vector $\vec{n}_r$.} 
The spin-dependent band structure of the
ferromagnets is described by the spin-dependent density of states (DOS), here
limiting ourselves to a \newer{flat band with
\begin{eqnarray}
  \nu_{r \sigma} (\omega) & = & \sum_k \delta (\omega - \varepsilon_{r k
  \sigma}) \text{ \ } = \text{ \ } \bar{\nu}_r (1 + \sigma n_r), 
  \label{eq:1dos}
\end{eqnarray}
for $| \omega | < W$ (half-bandwidth) and zero otherwise.} In Eq. (\ref{eq:1dos}), we
introduced the spin-averaged DOS $\bar{\nu}_r= (\nu_{r, \uparrow} +
\nu_{r, \downarrow}) / 2$ and the polarization $n_r =
\left| \vec{n}_r \right| = (\nu_{r, \uparrow} - \nu_{r, \downarrow}) /
(\nu_{r, \uparrow} + \nu_{r, \downarrow})$, which do not depend on the frequency $\omega$.
  \Figure{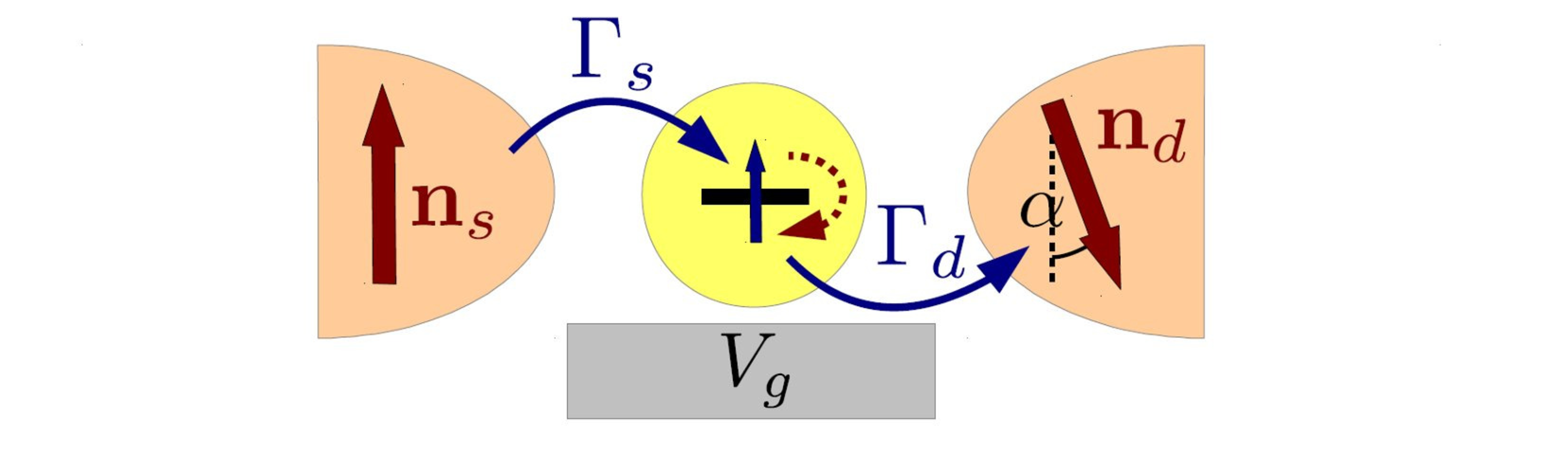}{Schematic
  setup of a QD spin valve, indicating the spin-precession resonance
  mechanism.\label{fig:model}}
\\
The breaking of the spin symmetry in the ferromagnets is expressed by
\begin{eqnarray}
  [ H_r, \hat{\vec{n}}_{r, \perp} \cdot \hat{\vec{S}}_r ] & \neq &
  0. 
\end{eqnarray}
Here, $\hat{\vec{n}}_{r, \perp} \cdot \hat{\vec{S}}_r$ is a component of the
spin operator $\op{\vec{S}}_r = \sum_{k \sigma \sigma'}^{} \text{}_r
{ \langle} \sigma { |} \hat{\vec{s}} {|} \sigma' { \rangle}_r c_{r k \sigma}^{\dag} c_{r k \sigma'}$
of ferromagnet $r$ along a unit vector $\hat{\vec{n}}_{r, \perp}$ that is
perpendicular to $\hat{\vec{n}}_r$. Note that for each ferromagnet, the axial
symmetry along its spontaneous magnetization direction, given by $\hat{\vec{n}}_r$,
remains intact: $[ H_r, \hat{\vec{n}}_r \cdot \hat{\vec{S}}_r ] =
0$. Importantly, the spin resonance relies on a {\tmem{complete}} breaking of
the spin symmetry by the ferromagnets, which means the full Hamiltonian does
not commute with \tmem{any} component \ $i = x, y, z$ of the total spin
operator $\op{\vec{S}}_{\tmop{tot}} = \op{\vec{S}} + \sum_r \op{\vec{S}}_r$,
that is,
\begin{eqnarray}
  [ H, \op{S}_{\tmop{tot}, i} ] & \neq & 0. 
\end{eqnarray}
This is achieved for noncollinearly polarized ferromagnets \new{with
polarizations $\vec{n}_s$ and $\vec{n}_d$ at an angle $\theta=\pi -
\alpha \neq 0,\pi$.}
Finally, the tunnel coupling Hamiltonian reads
\begin{eqnarray}
  \underset{}{\overset{}{H_T}}  & = & \sum_{r k \sigma} t_{r, \sigma \sigma'}
  d^{\dag}_{\sigma} c_{r k \sigma'} + \text{H.c.} \text{ \ } . 
\end{eqnarray}
Here, the tunneling amplitudes are assumed to be $k$- and therefore
energy independent as well as spin {\tmem{conserving}}, that is,
\begin{eqnarray}
  \underset{}{[ H_T, \op{\vec{S}}_{\tmop{tot}}]}  & = & 0. 
  \label{eq:tunspin}
\end{eqnarray}
However, since $d_{\sigma}^{\dag}$ and $c_{r k \sigma}$ may refer to different
spin quantization axes, the tunneling amplitudes
\new{
\begin{eqnarray}
  t_{r, \sigma \sigma'} & = &\langle
   \sigma { |} \sigma' \rangle_r \  t_r,  \label{eq:tssprime}
\end{eqnarray}
}incorporate an overlap factor of the spin states while the bare tunneling
amplitudes $t_r$ are spin-{\tmem{in}}dependent. They set the
spin-averaged tunneling rates by
$\Gamma_r = 2 \pi \bar{\nu}_r | t_r |^2$.
\subsection{Kinetic equations and charge current for infinite interaction
energy}\label{sec:notation}
The transport signatures of the QD spin valve are governed by the
nonequilibrium dynamics on the QD, described by its reduced density operator \
$\rho = \tmop{Tr}_{\tmop{res}} (\rho_{\tmop{tot}})$. Our reduced density
operator approach starts, as usual, from the von Neumann equation
$\dot{\rho}_{\tmop{tot}} = - i [H, \rho_{\tmop{tot}}]$ for the density \new{operator}
of the full system, $\rho_{\tmop{tot}}$. Eliminating the reservoir
degrees of freedom
results in the following kinetic equation for the reduced density operator of
the QD:
\begin{eqnarray}
  \dot{\rho} (t) & = & - i [H, \rho (t)] + W \rho (t) .  \label{eq:kineq}
\end{eqnarray}
Here, we have made an additional Markov approximation since we are interested
either \newer{in the stationary current} obtained from the stationary state satisfying $\dot{\rho}_{\tmop{st}} = 0$ (for
which it is irrelevant) or \newer{the time-dependent current} for which non-Markovian
corrections are of subordinate importance in our case as discussed in Appendix \ref{app:nonmarkov}. Thus, the effects due to the coupling to the leads are
incorporated through the zero-frequency kernel $W$.\\
To facilitate the analytical discussion of the QD spin dynamics, we express
Eq. (\ref{eq:kineq}) in terms of
coupled equations for the occupation probabilities $p_n$ for each of the
charge states $n = 0, 1, 2$, and the average spin $\vec{S} =
\tmop{Tr}_{\tmop{QD}} ( \op{\vec{S}} \rho )$. \new{The
equivalence of these two representations is
shown in Appendix \ref{app:liouville}.}
To keep all analytic expressions as simple as possible, we focus
first on the limit of $U \rightarrow \infty$, for which double occupancy
of the QD is suppressed (which implies $p_2=0$). In
this case, the kinetic
equations read
\begin{eqnarray}
  & \begin{array}{lll}
    \dot{p}_0 & = & - 2 \Gamma_0 p_0 + \Gamma_1 p_1 + 2 \vec{G}_{p S} \cdot
    \vec{S},\\
    \dot{\vec{S}} & = & + \vec{G}^0_{S p} p_0 - \tfrac{1}{2} \vec{G}^1_{S
    p} p_1 - \mathcal{R}_S \cdot \vec{S} - \vec{B} \times
    \vec{S},
  \end{array} &  \label{eq:spr-kineqdiscuss}
\end{eqnarray}
with $\dot{p}_0 = - \dot{p}_1$ due to probability conservation: $p_0 + p_1 =
1$. \newer{Equation (\ref{eq:spr-kineqdiscuss}) is the most general form of any
time-local quantum master equation for the QD system we study.} 
It extends common master
equation approaches for the occupation probabilities $p_n$ by including their
intense coupling (\New{through the vectors} $\vec{G}^0_{S p}$,$\vec{G}^1_{S p}$,$\vec{G}_{p S}$) to the
coherences  \cite{Leijnse08a,Koller10} of the degenerate spin states,
contained in the spin vector $\vec{S}$. Furthermore, the spin is subject to a torque
\New{corresponding to an effective exchange field $\vec{B}$.
This effective field arises from quantum fluctuations of QD electrons into the attached ferromagnets
and is the key factor in generating the spin resonance.}
Finally, the spin $\vec{S}$ is
subject to a spin decay, which is \New{described} by the symmetric
tensor $\mathcal{R}_S$. The
spin-\new{decay} tensor can become significantly anisotropic in the
Coulomb blockade regime due to cotunneling processes. This affects the width of the spin
resonance, which we discuss in Sec. \ref{sec:shape}. Extending usual master equations in the way described above is a necessity
for noncollinear spin valves, i.e., when the rotational symmetry is completely
broken (see Sec. \ref{sec:sprmodel}).\\
\newer{To compute \New{all coefficients} in Eq. (\ref{eq:spr-kineqdiscuss}), we
systematically expand the kernel in the tunneling rates $\Gamma_r = 2
\pi \bar{\nu}_r | t_r |^2$ using the real-time diagrammatic technique
\cite{Leijnse08a,Schoeller09a} and \new{we} include all leading-order $\Gamma$
and next-to-leading order $\Gamma^2$-terms. 
This has been done analytically, starting from a general
Liouville-space formulation of the real-time diagrammatic
approach \cite{Schoeller09a}.
The resulting expressions for the rates in the above quantum master equations
(\ref{eq:spr-kineqdiscuss}) are given in Appendix \ref{app:rates}. Our
results extend previous works in that we account for \emph{both} renormalization effects
due to the dynamics of coherences \emph{and}
next-to-leading order $\Gamma^2$ corrections. This
enables us to make \New{reliable} predictions about the spin resonance in the
Coulomb blockade regime.
\New{This development was motivated by Ref.\Cite{Baumgaertel11} where the spin
resonance was found at the flank of the single-electron tunneling peak but could not be tracked into the Coulomb blockade regime
because the quantum master equation used there included only leading order $\Gamma$ processes.}
The interested reader may find details on our
technical advances \New{and how they extend} previous works in Apps. \ref{app:rates}
and \ref{app:extension}.\\
\New{After} solving Eq. (\ref{eq:spr-kineqdiscuss}) for the occupation probabilities and
the average spin, one
can compute the average current from lead $r$ into the QD by
\begin{eqnarray}
  I_r & = & 2 \Gamma_{r,0} p_0 - \Gamma_{r,1} p_1 - 2 \vec{G}_{r, p S} \cdot
  \vec{S},  \label{eq:spr-current}
\end{eqnarray}
where the rates are given in Appendix \ref{app:rates}.
In Appendix \ref{app:crossover}, we explain how to solve
Eq. (\ref{eq:spr-kineqdiscuss}), which is actually a nontrivial task in the Coulomb
blockade regime because $O(\Gamma)$ contributions
can become smaller in magnitude than $O(\Gamma^2)$ contributions. With
the analytical results obtained in this way, we are able to gain a physical
understanding of the QD dynamics governing the spin resonance and they
are, moreover, used to derive approximation
formulas for the current near the resonance.\\
However, our analytical results are restricted to the limit $U
\rightarrow \infty$. To study the case of \emph{finite} charging energy $U$, we
use a computer code to evaluate the kernel $W$ numerically. The code
is based on the formulas given in Ref.\Cite{Leijnse08a}, which we
extended to account for couplings between diagonal
and nondiagonal \new{density operator} matrix elements to $O(\Gamma^2)$.
 To ensure that our perturbative approach is valid, we set for all
 plots \New{$\Gamma<T$} and we further checked that the numerically computed features
 scale at least as $O(\Gamma^2)$ when the tunnel coupling is
 lowered (see also comments in Appendix \ref{app:crossover}).
 \New{Thus, the predicted spin resonance in the Coulomb blockade regime can appear in an experiment also
 at elevated temperatures in the sense $T>\Gamma$, for which the Kondo effect is not present.}} \\
\section{Stationary-conductance resonance: characterization of quantum-dot
spin valves}\label{sec:resonance}
With our model and technique established, we \New{first} turn to the discussion of
the spin resonance in the stationary conductance. \New{Here, we} study the
generalization of Eq. (\ref{eq:spr-kineqdiscuss}) to finite $U$ and solve it
numerically unless stated otherwise. We first present the
most important features in the stationary conductance in
Sec. \ref{sec:overview} before we scrutinize the parameter dependence of the
resonance position and width in detail in the following sections. We
further expound that the nontrivial parameter dependence can be used to characterize QD spin valves in
an alternative way.
\subsection{Spin resonance in stability diagram}\label{sec:overview}
A main result of this \New{paper} is presented in Fig. \ref{fig:resonance}, which shows the stationary conductance for the setup
sketched in Fig. \ref{fig:resonance}(a) obtained from the
extension of Eqs. (\ref{eq:spr-kineqdiscuss}) and (\ref{eq:spr-current}) to finite
$U$. We find a sharp {\tmem{wiggle}} in the nonlinear conductance $d I / d
V_b$, i.e., a peak in the current plotted vs. $V_b$, which extends through the
entire Coulomb-blockade region. Notably, the resonance starts at the Coulomb
diamond edge, then bends towards the particle-hole symmetry point at $(V_b =
0, \ \varepsilon = - U / 2)$, where its magnitude vanishes, and then continues
point-symmetrically. We therefore focus our discussion first on the $V_g < U /
2$ part of Fig. \ref{fig:resonance} and chose the labels ``source'' and
``drain'' such that the lead with the larger spin-injection rate $\Gamma_r
n_r$ is the source for $V_b > 0$.
  \Figure{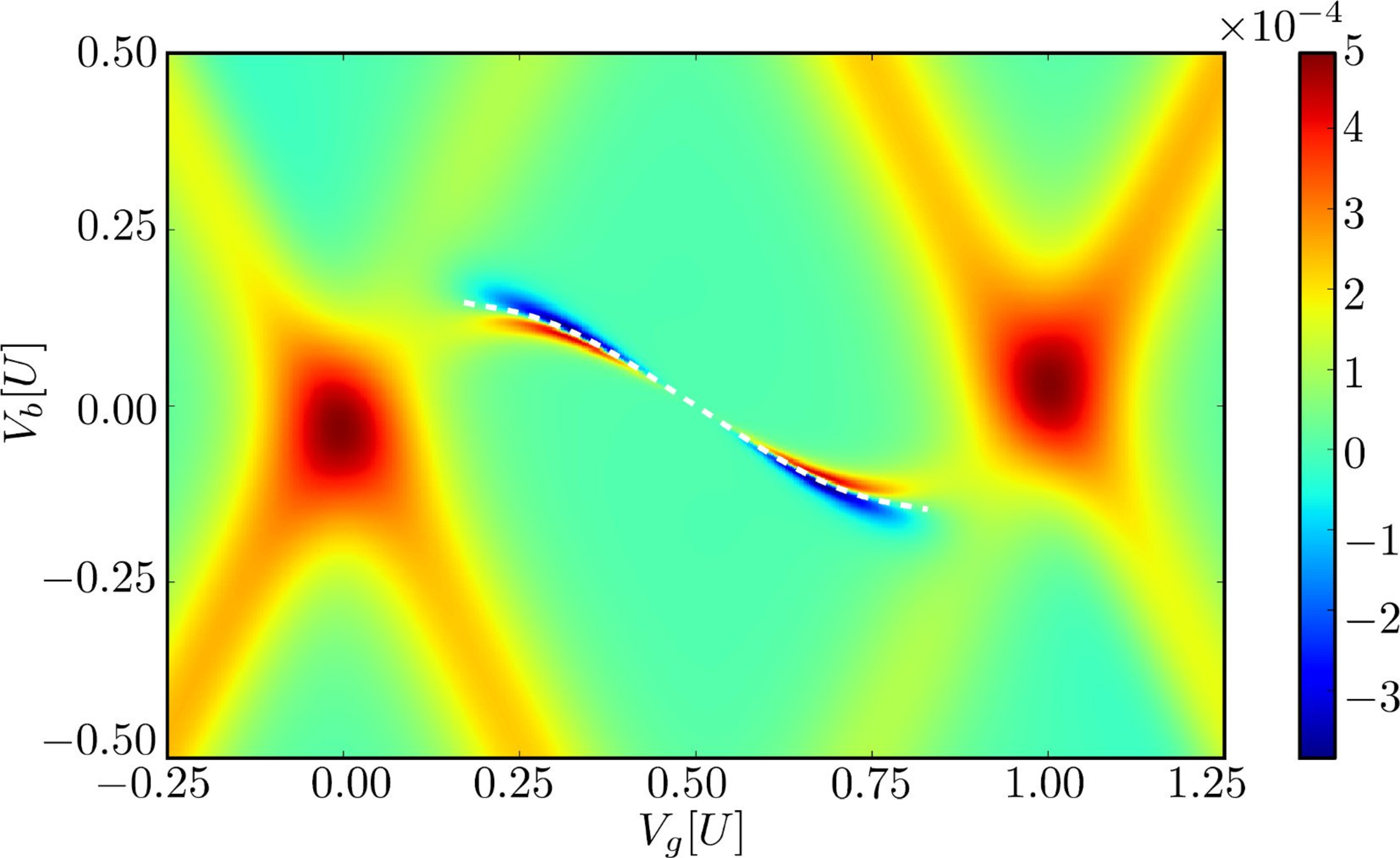}{Differential
  conductance $dI / dV_b$ for the setup shown in Fig. \ref{fig:model} for the current from the source into the QD for $\Gamma_s =
  2 \Gamma_d = 0.01 U$, $T = 0.05 U$, $W = 50 U$, $n_s = n_d = 0.99$, $\alpha
  = 0.01 \pi$. The white dashed curve follows from the resonance condition
  (\ref{eq:resonance}). Signatures in the conductance can already be found for
  \ $n_s, n_d \gtrsim 0.6$, and $\alpha < 0.4 \pi$ as discussed in
  Sec. \ref{sec:polangledep}; here we use larger
  polarizations and smaller $\alpha$ for illustrational
  purposes.\label{fig:resonance}}
\\
To understand the origin of the spin resonance, we note that the current
through the QD is largely suppressed for antiparallel polarizations by the
spin-valve effect: \new{Electrons} of spin-majority
type coming from the source get stuck in the QD because they are of spin
minority type for the drain. Thus, the tunneling rate for these electrons from the QD into
the drain is small. However, if the polarizations of the electrodes
are merely slightly {\tmem{noncollinear}}, the spin resonance appears in Fig.
\ref{fig:resonance}. The reason for this sharp resonance is that the drain
contribution to the exchange field, $\vec{B} = B_s  \hat{\vec{n}}_s + B_d 
\hat{\vec{n}}_d$, adds a component $B_{d, \perp} = B_d \sin \alpha$ that is
perpendicular to the source polarization $\vec{n}_s$, i.e., $\vec{B} = (B_s +
B_{d, \parallel})  \hat{\vec{n}}_s + B_{d, \perp} \hat{\vec{n}}_{\perp}$ with $B_{d,
\parallel} = B_d \cos \alpha$ [cf. Eq. (\ref{eq:dipappr}) below]. 
The seemingly innocuous component $B_{d,
\perp}$ causes a precession of the spin injected along $\vec{n}_s$ towards
$\vec{n}_d$. Consequently, the electron can easily leave the QD to the drain,
preventing an accumulation of spin \New{antiparallel} to the drain as expected from
prior works \cite{Koenig03,Braun04set}. 
We note that such a transverse component appears only because
of the noncollinearity of the ferromagnets' polarizations, i.e., the
complete breaking of the rotational symmetry. If the polarizations are
collinear, the exchange field is aligned along the common polarization
axis and therefore no spin precession is possible.\\
The spin-precession feature shown in Fig. \ref{fig:resonance} 
is unexpectedly sharp since the spin-valve effect is lifted {\tmem{only}} for
a specific bias voltage $V^{\ast}_b$. The reason is that the spin rotation is
effective only if the opening angle of the spin precession is large [cf. Fig.
\ref{fig:bfield}(c)(ii)]. Hence, the resonance appears when the total exchange
field component parallel to the source polarization $\vec{n}_s$ vanishes,
i.e., when the following scalar condition is satisfied:
\begin{eqnarray}
  & \begin{array}{lll}
    \vec{B} \cdot \hat{\vec{n}}_s \text{ \ } = \text{} B_s + B_{d, \parallel} & = &
    0.
  \end{array} &  \label{eq:resonance}
\end{eqnarray}
In contrast to usual resonance conditions, it incorporates two
{\tmem{vectors}}.\\
The resonance {\tmem{position}} can be predicted from the $O (\Gamma)$
approximation for the exchange field \cite{Braun04set},
\begin{eqnarray}
  \vec{B} & = & \sum_r \Gamma_r \vec{n}_r  [\phi_r (\varepsilon) - \phi_r
  (\varepsilon + U)],  \label{eq:dipappr}
\end{eqnarray}
with spin-polarization vector $\vec{n}_r$ pointing in the polarization
direction of the ferromagnet. Equation (\ref{eq:dipappr}) includes the
renormalization function
\begin{eqnarray}
  \phi_r (\varepsilon) & = & \int_{- W}^{+ W} \frac{d \omega}{\pi} \frac{f [ (\omega - \mu_r) / T]}{\omega - \varepsilon} \\
  & = & \frac{1}{\pi} \left[ - \tmop{Re} \psi \left( \frac{1}{2} + i
  \frac{\varepsilon - \mu_r}{2 \pi T} \right) + \log \left( \frac{W}{2 \pi T}
  \right) \right], \nonumber \label{eq:phi}
\end{eqnarray}
incorporating the digamma function $\psi$, the Fermi function $f (x) = 1 /
(e^x + 1)$\newer{, and electrochemical potentials $\mu_{s,d}=\pm V_b/2$}. Inserting Eq. (\ref{eq:dipappr}) into the resonance condition
(\ref{eq:resonance}) and solving for the resonant bias $V_b^{\ast}$ as a
function of $V_g$ yields the white dashed curve in
Fig. \ref{fig:resonance}.
\New{This simple physical idea thus nicely ties in with the results of our full numerical calculations as we further work out in Sec. \ref{sec:position}
and with our \new{analytical results based on the kinetic equations (\ref{eq:spr-kineqdiscuss}) for $U\rightarrow \infty$ in
Sec. \ref{sec:shape}. The full theory is, however, still needed for understanding the resonance peak height and shape.}}\\
Remarkably, for a given gate voltage $V_g$, the condition (\ref{eq:resonance})
is fulfilled only for one bias polarity when the electrodes are
asymmetrically coupled to the QD. \newer{This is one feature that can be used to rule out other
effects in experimental data, for example, those due to inelastic
cotunneling, which \new{typically} show signatures for both bias
polarities. Other \New{distinguishing} features are the peak height and width
as discussed in Sec. \ref{sec:shape}.\\ 
Here, we first focus \New{on} the explanation of the}
strong current rectification, which can be
attributed to the electrical tunability of the exchange field
{\tmem{direction}}: In Fig. \ref{fig:bfield}(a), we plot $B_s$, $B_{d,
\parallel}$, and their sum $B_{| |} = B_s + B_{d, | |}$ as function of the bias
$V_b$. For electrode $r$ the magnitude $B_r$ is maximal when $\mu_r =
\varepsilon$ or $\mu_r = \varepsilon + U$ and vanishes midway at $\mu_r =
\varepsilon + U / 2$ [marked in Fig. \ref{fig:bfield}(b) by (i) for $r = d$
and in (iii) for $r = s$]. In the vicinity of these points, the exchange field
$\vec{B}$ comes from only one electrode, pointing along $\vec{n}_s$ or
$\vec{n}_d$, see Fig. \ref{fig:bfield}(c)(i) and (iii), respectively. Here,
the spin precesses with a small opening angle and the spin transport stays
blocked. However, when tuning the bias between these two cancellation points,
the exchange field rotates [see Fig. \ref{fig:bfield}(c)(ii)] and the sum
$B_{\parallel}$ vanishes for a specific bias voltage $V_b^{\ast}$ and
polarity. This electric tunability illustrates that renormalization-induced
effective fields can intervene with the coherent evolution of two-level
systems in a controlled way to produce unexpected resonances as shown in Fig.
\ref{fig:resonance}.
  \Figure{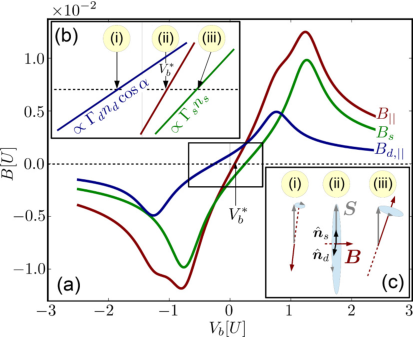}{Main panel (a) and sketched zoom-in
  (b): Exchange field component along $\vec{n}_s$ from the source electrode
  ($B_s$, green), the drain electrode ($B_{d, \parallel}$, blue), and their
  sum ($B_s + B_{d, \parallel}$, red) as a function of $V_b$ for $V_g = 0.375
  U$, with other parameters as in Fig. \ref{fig:resonance}. (c) Illustration
  of the spin precession (gray) for different directions of the exchange field
  (red), taken for different $V_b$ as indicated in (b). The opening angle is
  maximal for (ii) at $V_b = V_b^{\ast}$ where Eq. (\ref{eq:resonance})
  holds.\label{fig:bfield}}
\\
Figure \ref{fig:resonance} further clearly shows that the bias scale
$V_b^{\ast}$ does not match any obvious energy scale of the problem, attesting
to its nonspectral, vectorial nature. Depending on the gate voltage, it may
exceed $\Gamma$, $T$, and even approach a sizable fraction of $U$ \New{[cf. Fig. \ref{fig:limits1}(a)].} As we show
in Sec. \ref{sec:position}, the effect may be exploited to
characterize QD spin valves {\tmem{in situ}}.\\ 
\New{Similarly, additionally
attaching a superconductor to the QD, see Sec. \ref{sec:sc},}
\new{the spin-resonance position remains distinct from the energy scales set by the
Andreev bound states formed on the QD \cite{MartinRodero11}. The effect
of the Andreev bound states is to modify the exchange field $\vec{B}$
\cite{Sothmann10c}, which shifts the resonance position in the full
calculation notably as accurately predicted by the resonance condition
(\ref{eq:resonance}) when inserting the modified exchange field $\vec{B}$. This
is explained further in Sec. \ref{sec:sc}.}
\New{The above confirms that} Eq. (\ref{eq:resonance}) truly captures
the essence of the \New{spin resonance} under various
situations and \New{identifies a mechanism of a} highly voltage-dependent loss of magnetoresistance
for QD spin valves \New{that is active} already for small noncollinearity angles.\
\subsection{Experimental feasibility: Polarization
vectors}\label{sec:polangledep}
\New{In the above section, we used large polarizations $n_s = n_d = n = 0.99$
for illustrational purposes.
Achieving such high values is a central goal of spintronics,
\new{yet} currently presents a challenge.
\new{However, this large value was only used to make the resonance as clear
as possible in Fig. \ref{fig:resonance} but to observe our predicted
feature, this is actually not needed.}}
\New{In Fig. \ref{fig:andep}(a), we show} the
nonlinear conductance $d I / d V_b$ in the stationary state for \new{lower}
values of the polarization $n = n_s = n_d$. Clearly, already for polarizations
$n \gtrsim 0.6$ a discernible modification of the conductance can be seen.
\emph{In situ} polarizations as large as $n \sim 0.8$
\newer{have already been achieved} with half-metallic
electrodes \newer{in experiments} \cite{Hueso07}.
\\
\New{We also note that the small noncollinearity angle $\alpha = 0.01
\pi$ used in Fig. \ref{fig:resonance} shows that the assumption of perfect collinearity often made in theoretical analyses of spin-valve devices can lead to highly nongeneric results.}
\new{However,} the spin resonance is not limited to small noncollinearity
angles: Figure \ref{fig:andep}(b) shows the conductance in the vicinity of the resonance for
different noncollinearity angles $\alpha$ and one finds a region of
negative differential
conductance even for $\alpha$ as large as $0.4 \pi$. 
We conclude that it is
not essential to \New{have a} noncollinearity angle very precisely close to $\alpha
= 0$ \new{and extraordinary large polarizations $n \approx 1$} to see a resonance
feature in the stability diagram. \new{Large polarizations of $n \gtrsim
0.8$ as aimed at by efforts in spintronics
and
angles $\alpha \lesssim 0.2 \pi$ should be sufficient to \new{observe}
features of the spin resonance.}
Moreover, \New{Fig. \ref{fig:andep}(b) shows that the
resonance position changes as a function of the angle $\alpha$. This can be exploited} to measure the angle
$\alpha$ as we \New{discuss in the next section.}
  \Figure{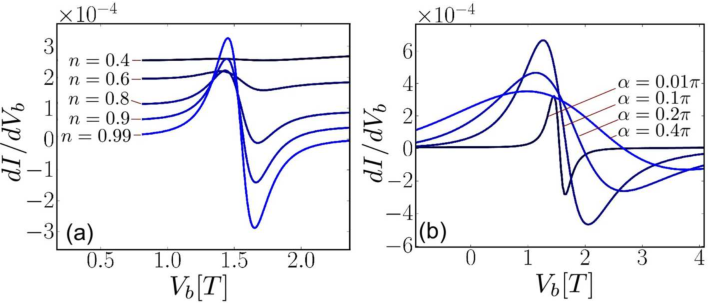}{Differential conductance $d I / d V_b$ as a
  function of bias voltage $V_b$ for gate voltage $V_g = 7.5 T$, varying (a)
  the polarization magnitude $n_s = n_d = n$ as indicated for fixed
  noncollinearity angle $\alpha = 0.01 \pi$ and (b) varying the angle $\alpha$
  as indicated for fixed $n_s = n_d = 0.99$. All other parameters are as in
  Fig. \ref{fig:resonance}(a).
  \label{fig:andep}}
\subsection{Extracting the spin-injection asymmetry from resonance
position}\label{sec:position}
To investigate the parameter dependence of the resonance position
systematically, \New{we introduce the energy level detuning from the symmetry point $\varepsilon=-U/2$,
\begin{align}
  \delta = U + 2 \varepsilon,
\end{align}
where the spin-resonance bias position goes through zero.
As shown in Appendix \ref{sec:spr-phs} using particle-hole symmetry,
it is sufficient to discuss only the case $\delta > 0$ and $V_b >0$
since the results obtained \new{are easily related to} those for negative values.
We thus limit our discussion here to the left half of the Coulomb diamond of the
stability diagram in Fig. \ref{fig:resonance}.
We} recast the resonance condition (\ref{eq:resonance}) as
\begin{eqnarray}
  \frac{a}{q} & = & 1,
  \label{eq:adivq}
\end{eqnarray}
with the asymmetry ratio of the spin-injection rates,
\begin{eqnarray}
  a & : = & \frac{\Gamma_s n_s}{\Gamma_d n_d \cos (\alpha)},
  \label{eq:r}
\end{eqnarray}
and electrically tunable ratio
\begin{eqnarray}
  q & : = & \frac{\phi_d (\varepsilon) - \phi_d (\varepsilon + U)}{\phi_s
  (\varepsilon) - \phi_s (\varepsilon + U)} .  \label{eq:q}
\end{eqnarray}
The above condition $a / q = 1$ has been used to generate the perfectly
matching white dashed curve in Fig. \ref{fig:resonance}(a)
by solving it for the resonant bias $V_b^{\ast}$ as function of $V_g$. Thus,
we find on a numerical basis that the $O (\Gamma)$ approximation for the
exchange field is sufficient to reliably predict the resonance position for
the full numerical calculation up to $O (\Gamma^2)$. Deep in the Coulomb
blockade regime when the distance of the electrochemical potentials from one
of the level positions is large, \newer{$\min_{r = s, d} [| \varepsilon - \mu_r |, | \varepsilon + U - \mu_r |] \gg T$}, the real part of the digamma
function can be approximated by a logarithm, that is, Re$\Psi (1 / 2 + ix)
\approx \ln |x|$. This leads to
\begin{eqnarray}
  q & \approx & \frac{\ln | (1 + \tilde{\delta} + \tilde{V}_b) / (1 -
  \tilde{\delta} - \tilde{V}_b) |}{\ln | (1 + \tilde{\delta} - \tilde{V}_b) /
  (1 - \tilde{\delta} + \tilde{V}_b) |} .  \label{eq:qlog}
\end{eqnarray}
Thus, the factor $q$ becomes independent of temperature and it exclusively
depends on the electrical parameters such as bias through the ratio
$\tilde{V}_b = V_b / U \geqslant 0$ and the gate voltage through the ratio
$\tilde{\delta} = 1 + 2 \varepsilon / U$. As a consequence, the resonance
feature is just rescaled inside the Coulomb diamond when the latter is made
larger by increasing the interaction energy $U$,
cf. Fig. \ref{fig:rconst}(d) below. We emphasize that the
nontrivial voltage dependence of the resonance position derives from the
drastic changes in the \New{\emph{direction}} of the exchange field vector $\vec{B}$,
rather than its magnitude.
  \Figure{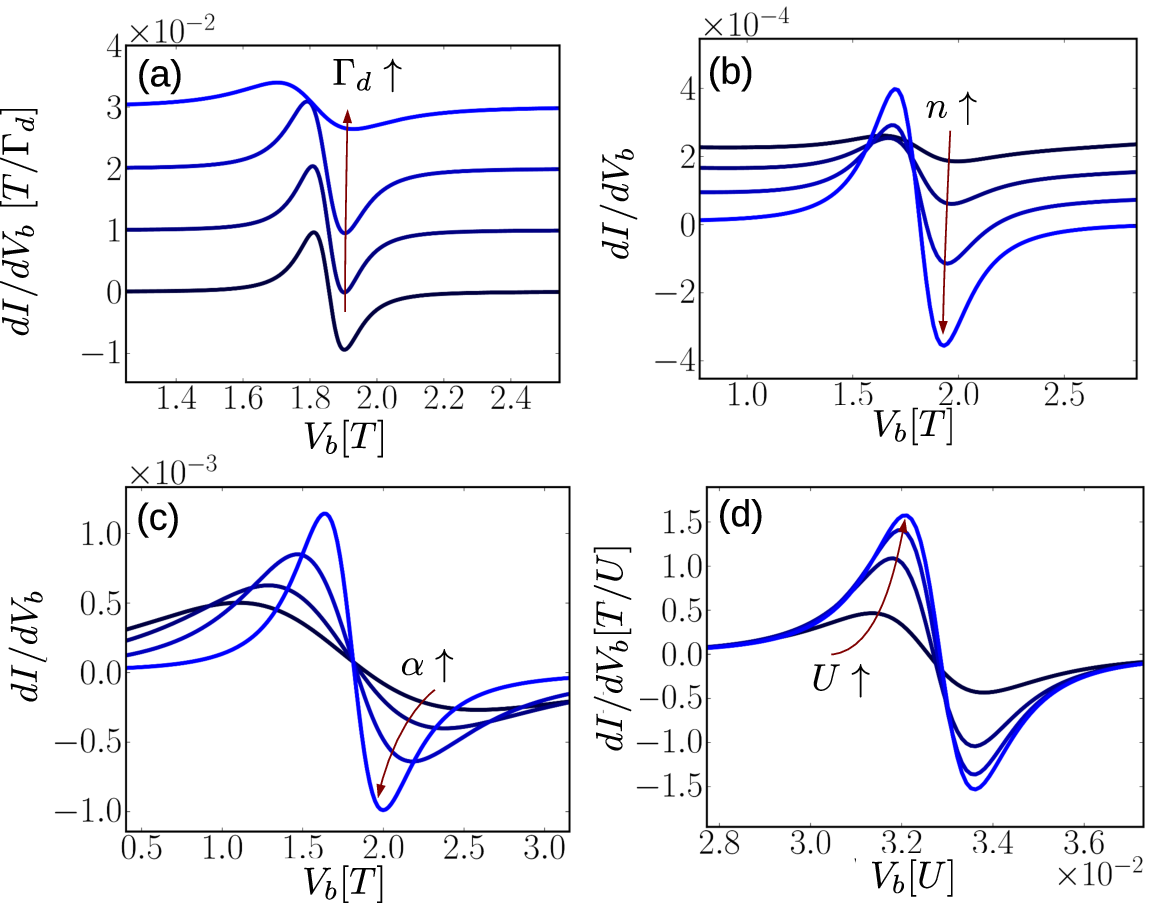}{Numerically
  computed differential conductance $d I / d V_b$ as a function of bias
  voltage $V_b$ when modifying several parameters but keeping the
  spin-injection asymmetry (\ref{eq:r}) fixed. (a) The tunnel
  couplings are varied as $\Gamma_d = \Gamma_s \text{/} (2 \cos (\alpha)) =
  10^{- 3}_{} T \ldots 10^{- 1} T$ in four equidistant steps, keeping $n_s =
  n_d = 0.99$ and $\alpha = 0.01 \pi$ fixed. The curves are
  vertically offset by $10^{-2}$ with respect to each other. (b) The polarization magnitudes
  are varied as $n_d = n_s  \text{/} \cos (\alpha) = 0.6 \ldots 0.99$ in four
  equidistant steps, keeping $\Gamma_d = \Gamma_s / 2 = 0.1 T$ and $\alpha =
  0.01 \pi$ constant. (c) The noncollinearity angle is varied as $\alpha =
  0.85 \pi \ldots 0.97 \pi$ in four equidistant steps, adjusting $\Gamma_d =
  \Gamma_s \text{/} ( 2 \sqrt{\cos (\alpha)} ) = 0.1 T$ and $n_d =
  n_s \text{/} \sqrt{\cos (\alpha)} = 0.99$. The parameters in (a)--(c)
  are chosen such that $a = 2$, the other parameters
  are $U = 20 T$, $V_g = 7 T$, and $W = 1000 T$. (d) The interaction energy is
  varied as $U = 40 T \ldots 100 T$ in four equidistant steps for $n_s = n_d =
  0.99$, $\Gamma_d = \Gamma_s \text{/} 2 = 0.1 T$, $\alpha = 0.01 \pi$, $V_g =
  0.45 U$, and $W = 1000 T$.\label{fig:rconst}}
\\
To substantiate the simple condition (\ref{eq:adivq}) further, we next show in
Fig. \ref{fig:rconst} \New{full} numerical results for the
resonance when changing various parameters in the setup such that the
asymmetry $a$ remains constant. According to our prediction from Eq.
(\ref{eq:adivq}), this leaves the resonant bias $V_b^{\ast}$ unchanged, which
is confirmed by Fig. \ref{fig:rconst}. For example, when
changing the tunnel couplings in Fig. \ref{fig:rconst}(a)
and the polarization in Fig. \ref{fig:rconst}(b), the
resonance \newer{width and height} are affected, but \New{the resonance bias \emph{position} indeed stays unaltered.}
 In Fig. \ref{fig:rconst}(c), we also
change the noncollinearity angle $\alpha$ while adapting both polarizations
and tunnel couplings to keep $a$ fixed. Finally, we increase in
Fig. \ref{fig:rconst}(d) the interaction energy $U$ and find
that the resonance condition (\ref{eq:adivq}) depends only on the ratios
$\tilde{V}_b = V_b \text{/} U$ and $\tilde{\delta} = 1 + 2 \varepsilon
\text{/} U$ of the voltages and the interaction energy for strong Coulomb
blockade conditions. \newer{By contrast}, the \New{width} of the resonance
 changes \New{significantly} because $U$ affects the spin-decoherence rates, see
Sec. \ref{sec:shape}.\\
We next \New{outline a simple procedure for determining} the asymmetry $a$ from an
experimentally measured stability diagram. Here, we use that the resonance
condition can be drastically simplified in the vicinity of the particle-hole
symmetry point. For $\tilde{\delta} \ll 1$, the condition $a / q = 1$ implies
that the resonant bias also satisfies $\tilde{V}_b^{\ast} \ll 1$. Then the
resonance position can be found by a linear expansion of the logarithms in Eq.
(\ref{eq:qlog}), which results in a {\tmem{linear}} dependence of the resonant
bias on the detuning,
\begin{eqnarray}
  \tilde{V}_b^{\ast} & = & \kappa (\alpha)  \tilde{\delta}, 
  \label{eq:linearcond}
\end{eqnarray}
with slope
\begin{eqnarray}
  \kappa (\alpha) & = & \frac{a (\alpha) - 1}{a (\alpha) + 1} . 
  \label{eq:kappa}
\end{eqnarray}
\new{The slope (\ref{eq:kappa}) becomes minimal in the limit $\alpha
\rightarrow 0$ (for $\alpha=0$ the spin resonance vanishes and the slope
cannot be measured). The slope increases quadratically with $\alpha$ as a simple
expansion of Eq. (\ref{eq:kappa}) for small $\alpha$ shows and reaches
$\kappa=1$ for $\alpha=\pi/2$.
Measuring the slope of the resonance position near the particle-hole
symmetry point in Fig. \ref{fig:resonance} allows one to directly
extract the spin-injection asymmetry $a(0)=\Gamma_s n_s \text{/} \Gamma_d
n_d$ and to measure the angle $\alpha$.
This can be achieved in two ways: First, if one has experimental access to this slope for a single,
accurately determined angle $\alpha$, one can directly determine $a(0)$. 
Alternatively, if one has continuous control
over $\alpha$ but the values for $\alpha$ are not known, one can
experimentally record pairs $[\alpha_i,\kappa(\alpha_i)]$ and use
Eq. (\ref{eq:kappa}) by inserting Eq. (\ref{eq:r}) as a fitting formula
with the single parameter $a(0)$. After these two possible ``calibration'' procedures, one can conversely extract the angle
$\alpha$ by measuring the slope.}
\new{All this} illustrates the usefulness of the novel spin resonance as alternative and simple route for
(partially) characterizing QD spin-valve setups \New{\emph{in-situ}}.\\

\subsection{Impact of proximal superconductor on resonance
position}\label{sec:sc}
To illustrate the broad applicability of our resonance concept, we study a
modification of model (\ref{eq:htot}) by adding a superconducting terminal
at electrochemical potential $\mu_{\sup} = 0$, tunnel coupled to the QD with
rate $\Gamma_{\sup}$, as sketched in Fig. \ref{fig:resonance_sc}(a). In the limit of infinite superconducting gap,
$\Delta \rightarrow \infty$, the effect of the superconductor can be
incorporated by adding a pairing term
\begin{eqnarray}
  H_P & = & - \tfrac{1}{2} \Gamma_{\sup}  (d^{\dag}_{\uparrow}
  d^{\dag}_{\downarrow} + d_{\downarrow} d_{\uparrow})\label{eq:pairing}
\end{eqnarray}
to the QD Hamiltonian (\ref{eq:hd}) \cite{Rozhkov00}.
  \Figure{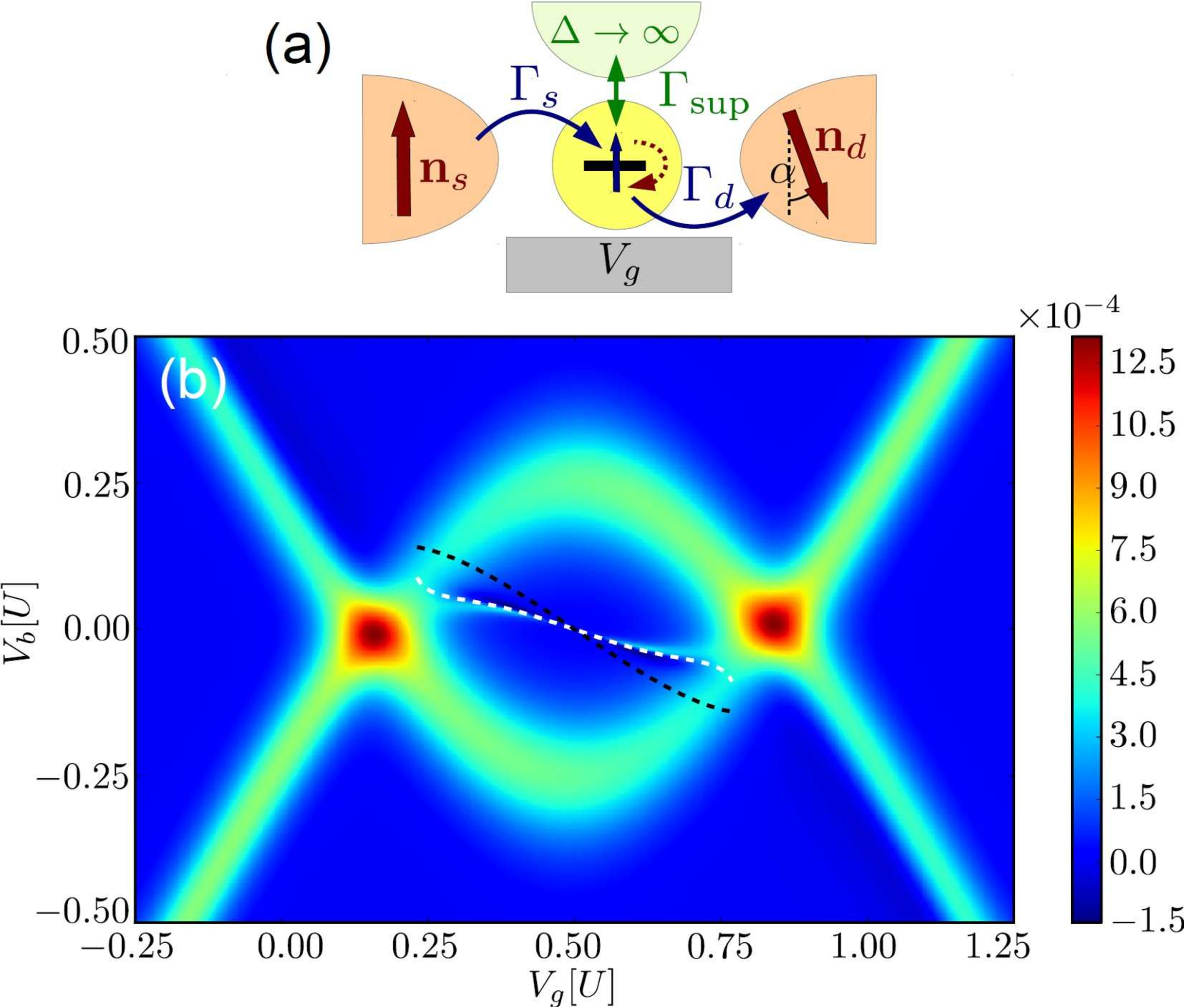}{(a)
  Modification of the quantum-dot spin valve depicted in
 Fig. \ref{fig:model}(a) including a superconducting terminal. (b) Differential
  conductance $dI / dV_b$ for setup (a) for the current from the source into
  the QD for $\Gamma_s = 2 \Gamma_d = 0.01 U$, \ $\Gamma_{\sup} = 0.75 U$, $T
  = 0.025 U$, $W = 50 U$, $n_s = n_d = 0.99$, $\alpha = 0.01 \pi$. The white
  (black) dashed curve follows from the resonance condition
  (\ref{eq:resonance}) including (excluding) the effect of the Andreev bound
  states. We excluded cotunneling from the calculations for (b). We comment on
  this in Appendix \ref{app:kineqsc}.\label{fig:resonance_sc}}
\\
In the presence of a superconductor, the dependence of the leading-order
exchange field on the electric parameters, contained in the ratio $q$ [Eq.
(\ref{eq:q})], is \New{modified: One has to replace Eq.
(\ref{eq:phi}) by  \cite{Sothmann10c}}
\begin{eqnarray}
  \phi_r (\varepsilon) & = & \sum_{\gamma \gamma' = \pm} \tfrac{\gamma'}{2
  \pi} \left( 1 + \tfrac{\gamma \delta}{2 \varepsilon_A} \right) \tmop{Re}
  \Psi \left( \tfrac{1}{2} + i \tfrac{\varepsilon_{r, \gamma' \gamma}}{2 \pi
  T} \right),  \label{eq:phimod}
\end{eqnarray}
with the modified energies $\varepsilon_{r, \gamma' \gamma} = \gamma' 
\frac{U}{2} + \gamma \varepsilon_A - \mu_r$ due to Andreev reflection
processes, incorporating the Andreev bound state energies $\varepsilon_A =
\frac{1}{2}  \sqrt{\delta^2 + \Gamma_{\sup}^2}$ for detuning $\delta = U + 2
\varepsilon$. In the limit of $\Gamma_{\sup} \rightarrow 0$, Eq.
(\ref{eq:phimod}) reduces to Eq. (\ref{eq:phi}). Solving the condition $a / q
= 1$ with $q$ modified through Eq. (\ref{eq:phimod}) for nonzero (zero)
$\Gamma_{\sup}$ gives the white (black) dashed curve in {
Fig. \ref{fig:resonance_sc}}(b). Clearly, the presence of the superconductor leads
to a significant shift of the resonance position.\\
Again approximating the real part of the digamma function deep in the Coulomb
blockade regime by a logarithmic expression, we find
\begin{eqnarray}
  q & = & \frac{\sum_{\gamma = \pm} \left( 1 + \tfrac{\gamma \tilde{\delta}}{2
  \tilde{\varepsilon}_A} \right) \ln \left| \frac{1 + 2 \gamma
  \tilde{\varepsilon}_A + \tilde{V}_b}{1 - 2 \gamma \tilde{\varepsilon}_A -
  \tilde{V}_b} \right|}{\sum_{\gamma = \pm} \left( 1 + \tfrac{\gamma
  \tilde{\delta}}{2 \tilde{\varepsilon}_A} \right) \ln \left| \frac{1 +
  2 \gamma
  \tilde{\varepsilon}_A - \tilde{V}_b}{1 - 2 \gamma \tilde{\varepsilon}_A +
  \tilde{V}_b} \right|},  \label{eq:qlogsc}
\end{eqnarray}
with $\tilde{\varepsilon}_A = \varepsilon_A / U$. The slope $\tilde{\kappa}$ of the linear resonance condition
$\tilde{V}_b^{\ast} = \tilde{\kappa}  \tilde{\delta}$, which is valid near the
particle-hole symmetry point, reads in this case:
\begin{eqnarray}
  \tilde{\kappa}\New{(\alpha)} & = & \kappa\New{(\alpha)} \ln \left| \frac{1 + \tilde{\Gamma}_{\sup}}{1 -
  \tilde{\Gamma}_{\sup}} \right| \frac{1 - \tilde{\Gamma}_{\sup}^2}{2
  \tilde{\Gamma}_{\sup}} \nonumber\\
  & = & \kappa\New{(\alpha)} (1 - \tilde{\Gamma}_{\sup}^2) + O (\tilde{\Gamma}_{\sup}^3), \label{eq:kappatilde}
\end{eqnarray}
with $\tilde{\Gamma}_{\sup} = \Gamma_{\sup} / U$ and $\kappa$ given by Eq.
(\ref{eq:kappa}). Hence, tuning the tunnel
coupling of a proximal superconductor does not only shift the single-electron tunneling resonance
positions in the stability diagram, but also suppresses the slope of the spin
resonance. This can be exploited to extract \New{the coupling to the superconductor $\tilde{\Gamma}_{\sup}$} in an
alternative way from the stability diagram: \new{If the tunnel coupling
$\Gamma_{\sup}$ can be effectively suppressed, which leads from Fig. \ref{fig:resonance_sc}(b) to
Fig. \ref{fig:resonance}, one can obtain
$\tilde{\Gamma}_{\sup}$  from Eq. (\ref{eq:kappatilde})
by inserting the experimentally measured values for
$\kappa(\alpha)$ and $\tilde{\kappa}(\alpha)$.}
This may be advantageous since
the broadening of the spin resonance can be much smaller than that of the single-electron tunneling
resonances, as demonstrated by Fig. \ref{fig:resonance_sc}(b).
\New{If one has additional control over the angle $\alpha$,} the broadening
of the spin resonance due to cotunneling processes\new{, which are not
included in Fig. \ref{fig:resonance_sc}(b),} can be
compensated by reducing \New{$\alpha$,} see Sec. \ref{sec:shape}).\\
\newer{We finally comment on our assumption of an infinite superconducting
gap $\Delta$. In experiments, the gap $\Delta$
can be $\sim$ 1 meV and
therefore on the order of typical \New{charging}
energies \cite{Franceschi10}. 
Hybrid
superconductor-ferromagnetic structures have also been realized with
somewhat smaller gaps of $\sim 100 \ \mu$eV \cite{Hofstetter10}.
However, as long as the bias $V_b$ is smaller
than $\Delta$ and the Andreev bound state
energies, real tunneling processes due to the superconductor are
strongly suppressed and renormalization effects due to quantum
fluctuations dominate.
 This expectation is underpinned by a recent theoretical
study \cite{Futterer13}, which considers corrections to the
\New{infinite}-gap approximation by expanding in $1/\Delta$. It turns out that
the main effect of the finite gap is to shift the Andreev bound state
energies rather than leading to modifications of the current. Therefore, we expect that
the form of our \New{resonance} condition (\ref{eq:resonance}) should be valid
for finite $\Delta$ when tuning
 close to the particle-hole symmetry point where the resonance
 appears for small bias.}\\
Our study thus illustrates a new,
fruitful and experimentally relevant interplay of superconductivity and spintronics.
Exploring the
situation of a finite superconducting gap $\Delta$ when $V_b \sim
\Delta$ is an interesting
open question that presents additional technical challenges beyond the scope of this
paper. For the rest of this paper, we return to the case when no
superconducting leads are present, i.e., $\Gamma_{\sup} = 0$.
\subsection{``Half-sided Coulomb diamond'' and zero-bias
peak}\label{sec:symmetric}
\New{As just illustrated by the superconducting hybrid setup,}
the spin resonance position sensitively reacts to modifications of the exchange field through the ratio $q$ regulating the dependence on voltages.
However, the \new{resonance position} can also be changed \new{by} the other factor in Eq. (\ref{eq:adivq}), the spin-injection asymmetry $a$. We
illustrate \New{this} for the two extreme cases
\New{leading to transport stability diagrams which would be puzzling if one were to experimentally obtain them without having further microscopic information:
For very large asymmetries, $a \gg 1$, the resonance becomes parallel to
Coulomb edges, forming a ``half-sided Coulomb diamond,''
whereas for negligible asymmetry, $a = 1$,  the resonance appears as a}
\new{zero-bias conductance peak.
Even though} \newer{the Kondo resonance and the zero-bias anomaly of
  Refs.{\Cite{Weymann05} and\Cite{Weymann05b}} also appear at zero bias,
  our spin resonance is clearly distinguished from these features as we explain below.}
  \Figure{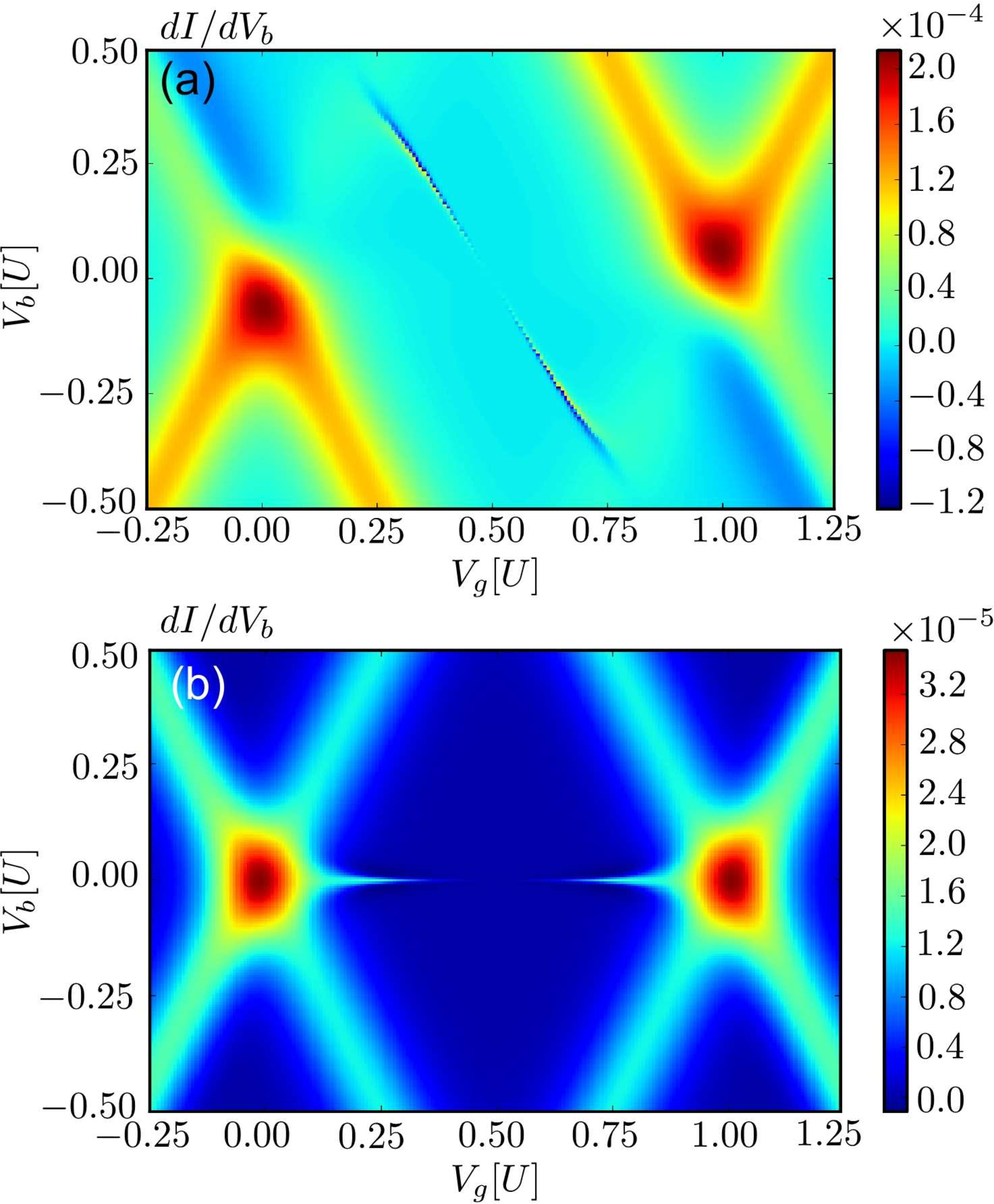}{Differential conductance $d I
  / d V_b$ as a function of gate voltage $V_g$ and bias voltage $V_b$. In (a),
  the spin-injection ratio is $a = 10$ with $\Gamma_s = 0.1 \text{/}
  \sqrt{\cos (\alpha)}$ and $\Gamma_d = 0.01 \text{/} \sqrt{\cos (\alpha)}$
  and in (b) the spin-injection ratio is $a = 1$ with $\Gamma_s = \Gamma_d =
  0.01 \sqrt{\cos (\alpha)} T$. All other parameters are as in
  Fig. \ref{fig:resonance}. 
   \label{fig:limits1}}
\\
We first note that the resonance position can appear in the entire voltage
range by changing $\kappa$ through the tunneling rates, polarizations, and the
angle $\alpha$, limited only by the condition
\begin{eqnarray}
  & 0 \text{ } \leqslant \text{ } \tilde{V}_b^{\ast} \text{ } \leqslant
  \text{ } \tilde{\delta}, & (\tilde{\delta} > 0)  \label{eq:restr}
\end{eqnarray}
if the electrode with the larger spin injection rate becomes the source for
$V_b > 0$ \new{and if $\alpha < \pi/2$ (which is needed for a sharp feature).} The restriction (\ref{eq:restr}) is readily proved from Eq.
(\ref{eq:adivq}): Since the asymmetry parameter $a = \Gamma_s n_s / [ \Gamma_d
n_d \cos (\alpha) ] \geqslant 1$, it follows that $q \geqslant 1$. The parameter
$q$ has a magnitude larger than 1 if the numerator in \ Eq. (\ref{eq:adivq})
is larger than that of the denominator, which implies $\tilde{V}^{\ast}_b
\geqslant 0$. For $q$ to be positive, one additionally has to demand
$\tilde{V}_b^{\ast} \leqslant \tilde{\delta}$ since $\tilde{\delta} \geqslant
0$. The analogous constraint in the other half of the Coulomb-blockade region,
\begin{eqnarray}
  & \tilde{\delta} \text{ } \leqslant \text{ } \tilde{V}_b^{\ast} \text{ }
  \leqslant \text{ } 0, & (\tilde{\delta} < 0), 
\end{eqnarray}
follows by similar arguments (see Appendix \ref{sec:spr-phs}).
\New{The above inequalities \New{turn into an equality} for the two extreme cases
illustrated in Fig. \ref{fig:limits1}.}
  \Figure{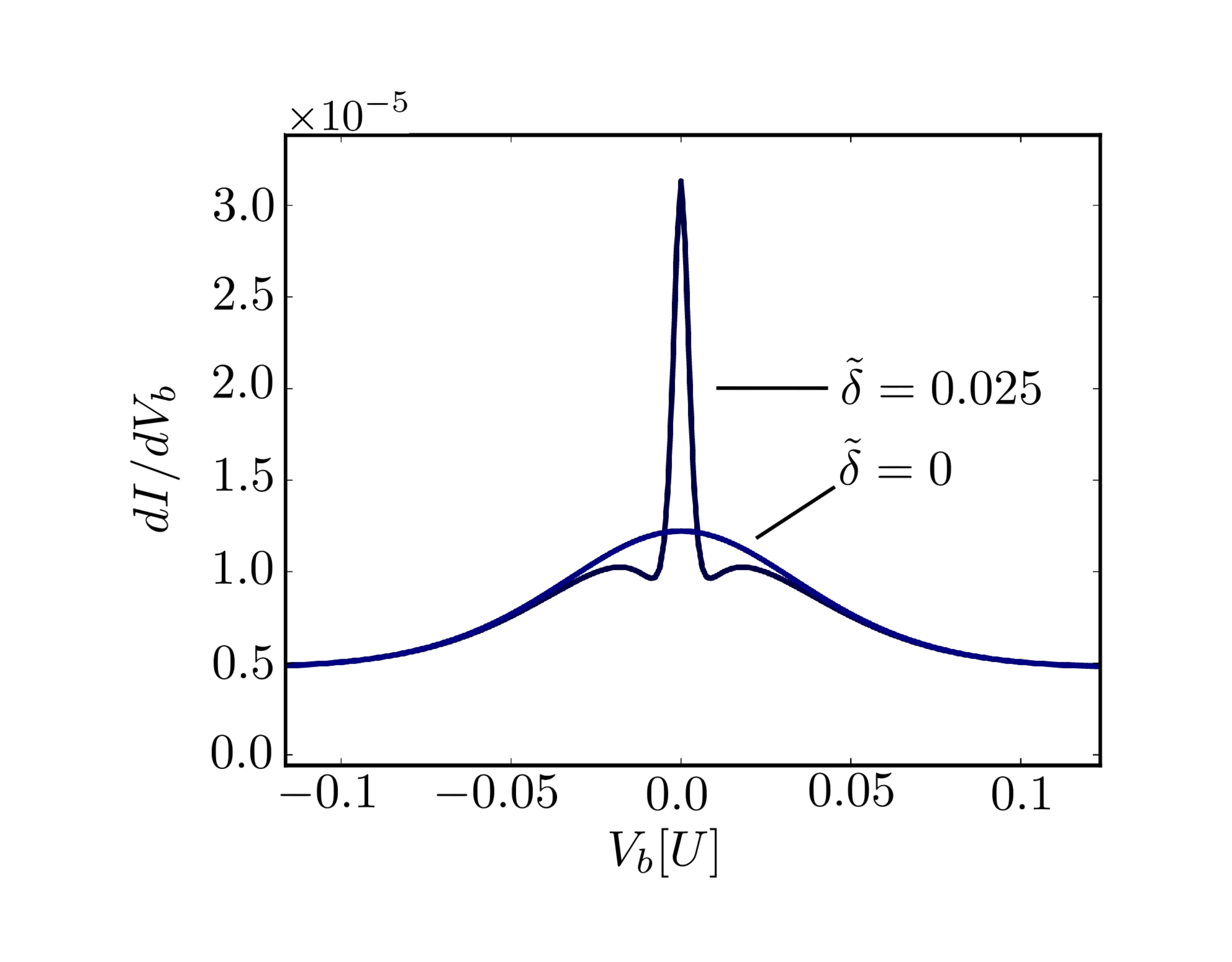}{
  Differential conductance $d
  I / d V_b$ as a function of bias voltage $V_b$ for $\Gamma_s = \Gamma_d =
  0.5 T$, $n_s = n_d = 0.99$, $\alpha = 0.01 \pi$, $U = 40 T$, and $W = 1000
  T$. For the gate voltage that restores the particle-hole symmetry,
  $\tilde{\delta} = 0$, the spin resonance is absent and the broad zero-bias
  anomaly of Refs.\Cite{Weymann05} and\Cite{Weymann05b} is visible. For $\tilde{\delta} = 0.025$, when the particle-hole
  symmetry is absent, the conductance profile for small bias is by contrast
  completely dominated by the spin resonance. We chose here rather larger
  tunnel couplings to compare with the above-cited references. We
  note that the conductance is shown there for strictly antiparallel
  polarizations, $\alpha = 0,$which has negligible impact \new{on the
  zero-bias anomaly} as compared to the
  case $\alpha = 0.01 \pi$ considered here. Note that the zero-bias
  anomaly does not appear in Fig. \ref{fig:limits1} because the tunnel
  couplings are smaller there, yet our spin resonance persists for these
  parameter values.
   \label{fig:limits2}}
\\
First, we show the
resonance in Fig. \ref{fig:limits1}(a) for strong asymmetry
$a \gg 1$. Here, the resonance position is at $\tilde{V}_b^{\ast} =
\tilde{\delta}$, i.e., parallel to the Coulomb diamond edges. Strikingly, the
resonance is much sharper than the \New{temperature-broadened} single-electron tunneling resonances
 because the width is not simply limited by temperature $T$ and tunnel coupling $\Gamma$ (see
Sec. \ref{sec:shape}).
\New{If one were to measure such a signature and would have no further microscopic information
 one would thus wonder why this feature is not thermally broadened,
 whereas the others can be demonstrated \new{to change with temperature.}}
\\
Second, \New{in Fig. \ref{fig:limits1}(b)} we show the resonance for perfect symmetry, $a = 1$,
\New{in which case it appears at $\tilde{V}_b^{\ast} = 0$ and only for an odd number of electrons on the QD.
This signature could in fact be mistaken for \new{features due to the Kondo
effect for electrodes with negligible polarization or the zero-bias anomaly discussed in
Refs.{\Cite{Weymann05} and\Cite{Weymann05b}}, which are otherwise very dissimilar.
One should note that the Kondo effect requires
strong tunnel couplings ($\Gamma/T \gg 1$), whereas the spin resonance also appears in the intermediate
coupling regime (still $\Gamma/T < 1$). Moreover, the spin resonance
disappears at the particle-hole symmetry point, while the Kondo effect
can remain at this
point. It can even appear \emph{only} at this point for strong, parallel spin polarizations of the electrodes \cite{Martinek03a,Martinek03b,Pasupathy04kondo,Hauptmann08} 
since the exchange field $\vec{B}=\vec{0}$ there (For the case of strong magnetizations of the electrodes this is no longer true \cite{Gaass11}).
For strong, antiparallel polarizations -- the configuration close to
\new{where} the spin resonance occurs -- it depends on the asymmetries of
the spin-injection rates whether the
Kondo effect emerges or not.\\
Furthermore,} for symmetric spin-injection rates, one should also not mistake the spin resonance for the zero-bias anomaly studied in Refs.{\Cite{Weymann05}
and\Cite{Weymann05b}}, caused by the interplay of the voltage dependence of the cotunneling spin-flip rates with the spin-valve effect.
Both effects can in fact} appear together and, as we demonstrate
in Fig. \ref{fig:limits2}, the spin resonance may be even
much larger and sharper than the zero-bias anomaly. However, it depends on the
choice of the parameters which of the two is more pronounced: For example,
while the width of the zero-bias anomaly is set by temperature, the width of
the spin resonance \New{is independent of $T$ and} determined
\New{instead} by the angle $\alpha$ and a combination of the spin-decay
rates and the exchange field, \newer{which depends strongly on the applied gate voltage}
(see Sec. \ref{sec:shape}). Moreover, in contrast to the
spin resonance, the zero-bias anomaly persists at the particle-hole symmetry \New{point}
and for \New{strictly} antiparallel lead polarizations.
\\
\New{The above  illustrates that our spin
resonance is really an independent conductance feature, distinct from other features and can
moreover be identified unambiguously in an experiment.}
\subsection{Extracting the anisotropy of the spin-decay tensor from the resonance
shape}\label{sec:shape}
\New{Besides the resonance position we have focused on so far,
the \New{resonance shape}} provides additional valuable information about the QD
spin-valve: In particular, one can extract information about the anisotropy of the spin-decay rates,
that is, the spin-relaxation rate $\Gamma_{| |} = \op{\vec{n}}_s \cdot
\mathcal{\mathcal{R}}_S \cdot \op{\vec{n}}_s$ and the
spin-dephasing rate $\Gamma_{\perp} = \op{\vec{n}}_{\perp} \cdot
\mathcal{\mathcal{R}}_S \cdot \op{\vec{n}}_{\perp}$, where
$\op{\vec{n}}_{\perp}$ is a unit vector perpendicular to $\op{\vec{n}}_s$. In
contrast to the position, the shape is significantly influenced by cotunneling
corrections \New{and crucially relies on the technical developments we report.}\\
\New{To illustrate this, we now restrict our attention} to voltages near the resonance (such that
$B_{\parallel} / B_{\perp} \lesssim 1$) in the limit of strong Coulomb
blockade ($V_b / 2 \varepsilon \ll 1$ and $U \rightarrow \infty$), small
noncollinearity angles ($\alpha \ll 1$), symmetric polarization magnitudes
$n_s = n_d = n$, and small spin injection asymmetry ($\kappa \ll 1$). In this
case, the stationary current (\ref{eq:spr-current}), $I = I_s = - I_d$,
flowing through the QD can be approximated by
\begin{eqnarray}
  I^{\tmop{appr}} & = & I_0  [1 - A (1 - M)],
  \label{eq:iapprox}
\end{eqnarray}
with
\begin{eqnarray}
  I_0 & = & \frac{\sum_{r, \tau} r (- 1)^{\tau} \Gamma_{r, \tau}
  \Gamma_{\bar{r}, \bar{\tau}}}{2 \Gamma_0 + \Gamma_1},  \label{eq:i0}\\
  A & = & 2 \frac{\sum_{r, \tau} r (- 1)^{\tau} \left( \vec{G}_{r,p
  S} \cdot \op{\vec{n}}_s \right) \left( \vec{G}^{\tau}_{S p} \cdot
  \op{\vec{n}}_s \right) \Gamma_{\bar{r}, \bar{\tau}}}{ \sum_{r, \tau} (-
  1)^{\tau} \text{ } \Gamma_{| |} \Gamma_{r, \tau} \Gamma_{\bar{r},
  \bar{\tau}}}, \\
  M & = & \frac{M_0}{1 + [(a \text{/} q - 1) \text{/} H]^2}, 
\end{eqnarray}
where all rates are defined in Appendix \ref{app:rates}. Here,
$\tau$ takes the values $0$ and $1$, $\bar{\tau} := 1 - \tau$, and the
factor $r$ in the above sums takes the value $r = +$ ($r = -$) for $r = s$ ($r
= d$) \New{and $\bar{r}=-r$.} Finally, we introduced the abbreviations
\begin{eqnarray}
  M_0 & = & \frac{1}{1 + (\Gamma_{\parallel} \Gamma_{\perp}) / B_{\perp}^2}, 
  \label{eq:m0}\\
  H & = & \alpha \sqrt{\frac{\Gamma_{\perp}}{\Gamma_{\parallel} M_0}} . 
\end{eqnarray}
Equation (\ref{eq:iapprox}) can be interpreted as follows: The value $I_0$ is
the current obtained when ignoring the spin accumulation, that is, forcing
\New{``by hand''} $\vec{S} = 0$ in the kinetic equations (\ref{eq:spr-kineqdiscuss}). Note that $I_0$  does
not coincide with the current for zero polarization \new{since the
charge-relaxation rates (\ref{eq:cot}) also depend on the
polarizations}. The actual nonzero spin
accumulation $\vec{S} \neq 0$ on the QD acts back on the charge dynamics, thus
suppressing the current to a fraction $1 - A < 1$ of the current $I_0$.
However, for any nonzero $\alpha$ the exchange-field induced spin precession
can \New{in turn} suppress this spin-valve effect. This is captured by the factor $1 - M$,
where $M$ is a Lorentzian function in the parameter $a / q - 1$ with
intensity $M_0$ and width $H$. The current becomes maximal at $a / q = 1$,
which is the resonance condition (\ref{eq:adivq}).\\
The peak value of the current resonance depends on two competing influences of
the cotunneling contributions to the current: On the one hand, they increase
the maximally achievable current $I_0$ by providing additional tunneling
processes, but on the other hand they enhance the spin decay, which limits the
effectiveness of the spin precession by suppressing $M_0$ \New{and thereby $M$}. The decisive
parameter that controls the current peak value is the ratio
\begin{eqnarray}
  b & : = & | B_{\perp} | \text{/} \sqrt{\Gamma_{\perp} \Gamma_{| |}} 
  \label{eq:bratio}
\end{eqnarray}
of the perpendicular exchange field component and the spin-decay rates.
Notably, the spin resonance appears both for (i) the strongly underdamped case
$b \gg 1$ and for (ii) the critically damped case $b \sim 1$, while it
disappears for (iii) the strongly overdamped case when $b \ll 1$, \New{where $M_0
\to 0$} and therefore $M$ has negligible impact on the current. The
``optimal'' value for a maximal current enhancement is given for $b \approx
1$. However, even for $b < 1$ but not yet $b \ll 1$, the spin precession can
still significantly enhance the current to produce a sharp feature in the
conductance \New{as in Fig. \ref{fig:resonance}}. \new{Therefore, the occurrence of the spin resonance in the
{\tmem{stationary}} conductance is \emph{not} yet evidence of an underdamped spin
precession.} By contrast, the pulsing scheme discussed \New{below} in
Sec. \ref{sec:pulsing} is able to unambiguously demonstrate  underdamped spin precession.
\\
Before discussing the time-dependent results, we first compare the stationary features in \New{the cases (i) and (ii)} and moreover explain
how they can be exploited to extract the spin-decay \new{properties} from electron transport measurements.
A salient finding of this scheme is that the electrical
tunability of the exchange field allows for an {\tmem{all-electric}} probing
of the anisotropic spin-decay tensor $\mathcal{R}_S$ {\tmem{in-situ}}. This
scheme resembles that of Ref.{\Cite{Braun05a}} where the
interplay of the exchange field with an external perpendicular magnetic field
was used to extract the spin-dephasing rate. Here, one utilizes the built-in
exchange field instead.
  \Figure{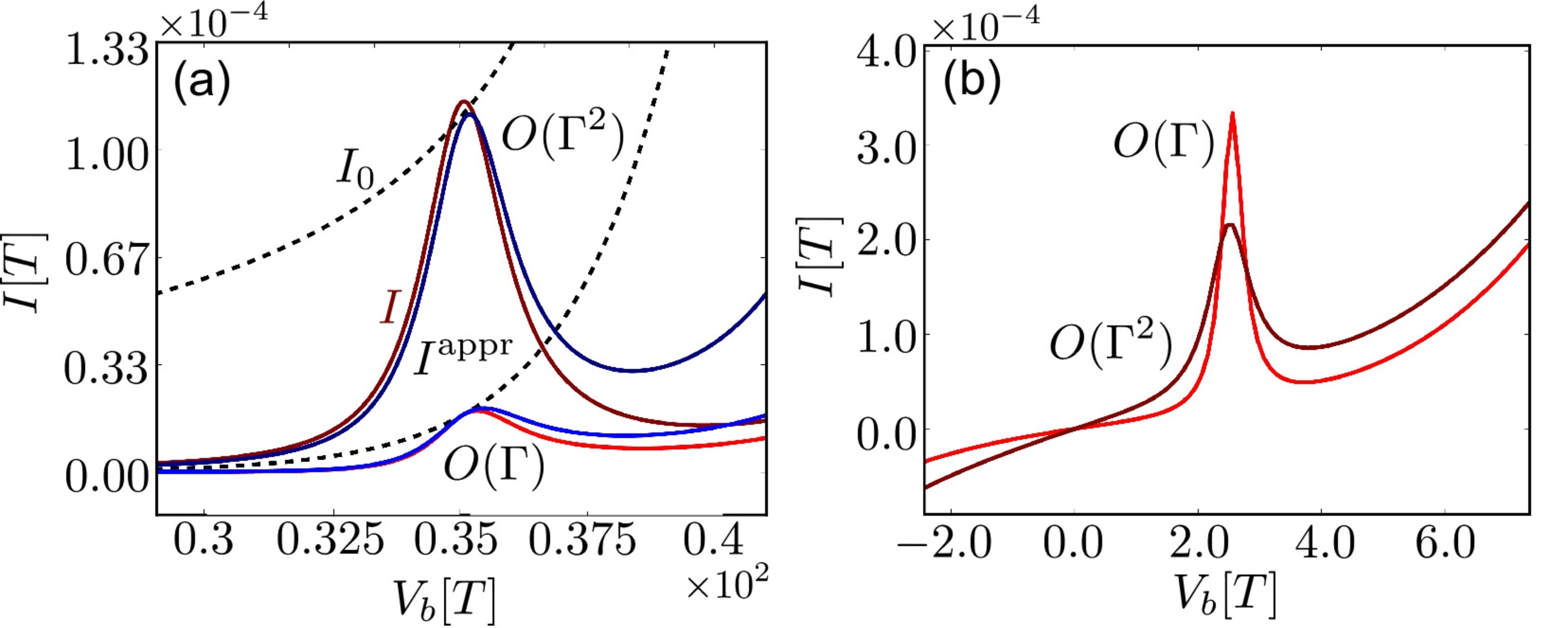}{  Stationary
  current as a function of bias voltage $V_b$ up to $O (\Gamma)$ and $O(\Gamma^2)$ as indicated.
(a)
  Strongly underdamped case ($b \gg 1$) in the large-$U$ limit. The
  current is computed first numerically from the extension of Eq.
  (\ref{eq:spr-kineqdiscuss}) to the finite-$U$ case for $U = 1000 \text{ } T \text{/}
  3$ (red, denoted by $I$) and approximated by formula (\ref{eq:iapprox}) in
  the limit of $U \rightarrow \infty$ (blue, denoted by $I^{\tmop{appr}}$). We
  also show the current (\ref{eq:i0}) for zero spin accumulation (dashed
  black, denoted by $I_0$). The other parameters are $V_g = 75 \text{ } T
  \text{/} 3$, $\Gamma_s = 2 \Gamma_d = 0.2 \text{ } T \text{/} 3$, $n_s = n_d
  = 0.99$, $\alpha = 0.01 \pi$, and $W = 1000 \text{ } T / 3$. This choice of
  parameters implies \new{$b \gtrsim 5$} [given by Eq. (\ref{eq:bratio})] for both $O
  (\Gamma)$ and $O (\Gamma^2)$ at the resonance. The approximated current and
  the numerically computed current match well but not perfectly. The main
  reason for the deviation is that the resonance does not appear here under
  strong Coulomb blockade conditions, as required for Eq. (\ref{eq:iapprox})
  to be \New{strictly} valid. These conditions are met in
  Fig. \ref{fig:timedep}(b) below, where approximation and numerical solution match
  perfectly. However, if we go deeper into the Coulomb blockade regime here,
  the resonance disappears in $O (\Gamma)$, cf. Ref.{\Cite{Baumgaertel11}}. Therefore, to make a comparison between the $O
  (\Gamma)$ and $O (\Gamma^2)$ current, we considered the resonance closer to
  the single-electron tunneling regime. (b) Critically damped case ($b \approx 1$) in the finite-$U$
  case: Stationary current up to $O (\Gamma)$ and $O (\Gamma^2)$ as a function
  of bias voltage $V_b$ numerically calculated from Eq.
  (\ref{eq:spr-kineqdiscuss}) for $V_g = 5 T$ with all other parameters as in
  Fig. \ref{fig:resonance}, implying $b \approx 0.5$ for $O
  (\Gamma)$, $b \approx 0.2$ for $O (\Gamma^2)$. Note that our approximation
  formula (\ref{eq:iapprox}) cannot be applied for \New{the} finite-$U$ case employed
  here.\label{fig:shape}}
\\
\New{(i) \emph{Underdamped regime} ($b \gg 1$).
In this regime,} the current is restored to the full
value $I_0$ at resonance ($a / q = 1$) since $M_0 \approx 1$. This is
illustrated in Fig. \ref{fig:shape}(a), in which we plot the
current numerically obtained from Eq. (\ref{eq:spr-kineqdiscuss}) extended to
finite $U$ and the approximation formula (\ref{eq:iapprox}). Both are close to
the value of $I_0$ (black dashed line) at the resonance. Both agree well, but
not perfectly, as we explain further in the caption of {
Fig. \ref{fig:shape}}. The resonance width,
\begin{eqnarray}
  H & \approx & \alpha \sqrt{\frac{\Gamma_{\perp}}{\Gamma_{\parallel}}}, 
\end{eqnarray}
directly yields the anisotropy of the spin-decay tensor, $\Gamma_{\perp} /
\Gamma_{\parallel}$, when the angle $\alpha$ is known. To extract
$\Gamma_{\perp} / \Gamma_{\parallel}$ from experimental data, one first
determines the spin-injection asymmetry $a$ from the resonance position, as
described in Sec. \ref{sec:position}. One then fits Eq.
(\ref{eq:iapprox}) to gate or bias traces of the current peak, expressing $a /
q - 1$ with the help of Eq. (\ref{eq:qlog}) as a function of bias and gate
voltage. In the resulting expression, the functions $I_0$, $H$, and $A$
appear. For fitting to experimental data, we suggest to treat these slowly
varying functions as constant fitting parameters near the resonance.\\
\\
\New{(ii) \emph{Critical damping} ($b \approx 1$).
When} the spin-decay rate
is comparable to the spin-precession rate, the current peak value is not
completely restored to $I_0$ as $M_0$ reaches only a fraction of 1. Here, the
spin decay limits the maximally achievable rotation angle for the QD spin
before it decays or tunnels out. This is visible in {
Fig. \ref{fig:shape}}(b), where the peak current may become {\tmem{smaller}} in $O
(\Gamma^2)$ as compared to that in $O (\Gamma)$, where the spin decay is much
slower. Furthermore, cotunneling corrections affect the width $H$ more
strongly than in the strongly underdamped regime: Here, the width is not
exclusively determined by the ratio $\Gamma_{\perp} / \Gamma_{| |}$ but also
incorporates $b$, which differs depending on whether cotunneling corrections
are included or not. 
This illustrates that -- in contrast to the resonance
position -- \New{for the accurate prediction of the resonance shape the next-to-leading order corrections are indispensable.} The pronounced
sensitivity of the resonance to cotunneling processes in the critically damped
limit $b \approx 1$ is also interesting for the characterization of the QD
spin valve: Once $B_{\perp}$ is determined, e.g., from the pulsing scheme (see
Sec. \ref{sec:larmor}), we may again use Eq.
(\ref{eq:iapprox}) as fitting formula, taking $M_0$ now as an additional
fitting parameter. One may then extract the spin relaxation rate
$\Gamma_{\parallel}$ and the dephasing rate $\Gamma_{\perp}$
{\tmem{individually}} by combining the results for $H$ and $M_0$.\\
\section{Gate-pulsing scheme: all-electric single-spin
operations}\label{sec:larmor}

In principle, the transport-induced spin decoherence time $\sim U / \Gamma^2$
can be made comparable or longer than experimentally measured spin-dephasing
times due to other mechanisms (see Sec. \ref{sec:estimate}) \new{by
reducing the tunneling rates}. Hence, multiple revolutions of an individual QD spin are
feasible. Probing this underdamped spin precession requires time-resolved
measurements. \new{At first sight, it may seem challenging to utilize our
transport setup for spin detection:
Many spin-to-charge conversion readout schemes rely on a large energy
splitting $B\gg T$ between the two spin states allowing the QD electron to
leave into an attached electrode only if it has one type of
spin. However, as there is no discernible spin splitting in our case,
such an energy-selective readout scheme \cite{Hanson07rev} is not
applicable here. We therefore suggest to employ a tunneling-rate selective
readout \cite{Hanson07rev}, which is naturally provided by the strongly
spin-polarized ferromagnets in our setup. As we predict in
Sec. \ref{sec:pulsing}, this only requires the adaptation of an experimentally
well-developed pulsing scheme  \cite{Fujisawa03rev}. Using this scheme,
underdamped oscillations in the \New{\emph{time-averaged current}} can
be probed as a function of the pulsing
duration.} To optimize the contrast in the average current oscillations, the
pulsing durations have to be chosen appropriately as we explain in
Sec. \ref{sec:optimal}.\\
\new{In contrast to other transport transport features in
the Coulomb blockade regime such as the Kondo effect or the zero-bias
anomaly discussed in Refs.{\Cite{Weymann05} and\Cite{Weymann05b}}, the spin resonance does not
destroy the coherence of the QD spin. This is an advantage as it allows} all-electric spin control to
be accomplished even without the need of an external magnetic field or
spin-orbit interaction. 
\New{Only the basic tool of spintronics is required: large polarizations of the ferromagnets.}
\subsection{Probing underdamped spin precession from average
current}\label{sec:pulsing}
The procedure of the simple pulsing \New{scheme  is} sketched in Fig.
\ref{fig:timedep}(a): \New{At fixed bias voltage $V_b$, one} repeatedly applies a rectangular voltage pulse to
the gate electrode, switching from $V_g^0$ to $V_g$ for a time duration
$\tau$, and then back to $V_g^0$ for a time duration $\tau^0$.
\New{Figure \ref{fig:timedep}(b) shows the stationary current as
function of $V_g$, exhibiting the spin resonance.}
We suggest to probe the time-averaged current over many pulses,
\begin{eqnarray}
  \bar{I} & = & \int_0^t \frac{d t'}{t} I_s (t') \text{ \ \ } (t \gg
  \tau, \tau^0),  \label{eq:avcurr}
\end{eqnarray}
varying the time duration $\tau$. Figures \ref{fig:timedep}(c)--\ref{fig:timedep}(e) \new{illustrate} that
the time-averaged current oscillates as a function of $\tau$ with a period
given by $2 \pi / | \vec{B} |$, which coincides with the period of the plotted
spin oscillations. Thus, one can extract the magnitude of the exchange field
$| \vec{B} |$ at $(V_b, V_g)$. The oscillations can be physically understood
as follows: By switching from the spin-valve blocked reference voltage $V_g^0$
[with field $\vec{B}^0$ nearly collinear with $\vec{n}_s$, cf. \New{panel (i) in Fig.
\ref{fig:timedep}(a)]} to a voltage $V_g$ where the exchange field $\vec{B}$
precesses the injected spin, the electron is more probable to escape upon
return to $V_g^0$ provided the duration $\tau$ matches a half-integer multiple
of the precession time $\tau_P = 2 \pi / | \vec{B} |$.
  \Figure{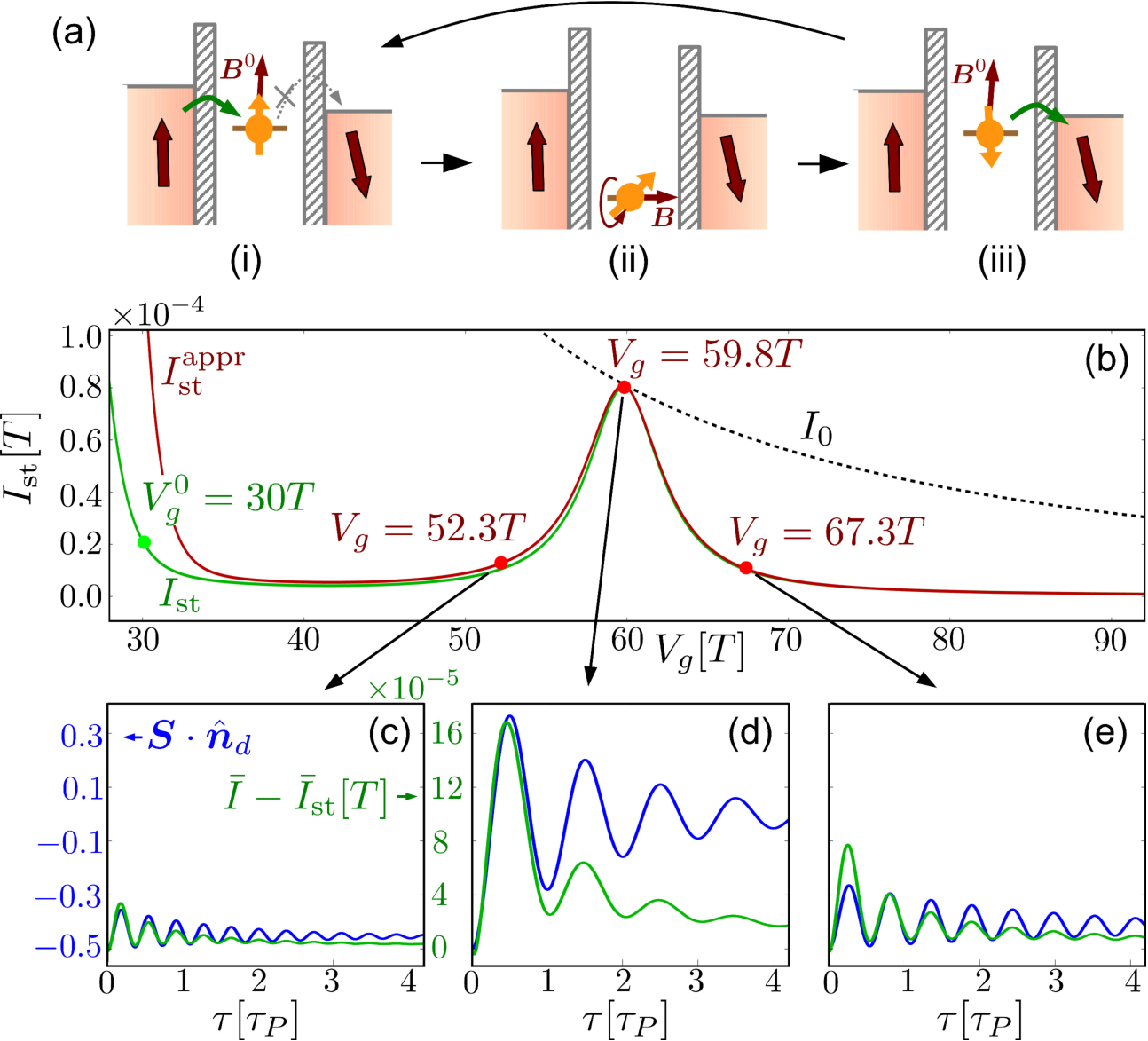}{(a) Schematics of the pulsing scheme. (b) Stationary current as
  function of $V_g$, obtained by solving Eq. (\ref{eq:spr-kineqdiscuss}) exactly
  ($I_{\tmop{st}}$, green), by neglecting the spin accumulation, i.e., forcing
  $\vec{S} = 0$ ($I_0$, dashed black), and by taking the approximation
  (\ref{eq:iapprox}) near resonance ($I^{\tmop{appr}}_{\tmop{st}}$, red), see
  Sec. \ref{sec:shape}. (c)--(e) Average current \New{$\bar{I} =
  \int_0^t (d t' / t) I_s (t')$} $(t \gg \tau, \tau^0)$ (green curves) as
  a function of $\tau$ for three different $V_g$ as indicated and for fixed
  $\tau^0 = 2 \cdot 10^3 / T = 0.46 \tau_P$, and $V_g^0 = 30 T$. The times
  $\tau^0$ and $\tau$ are given in units of the precession period at
  resonance, $\tau_P \approx 4.7 \cdot 10^3 / T$. The current is offset by
  $\bar{I}_{\tmop{st}}$, the current that would flow if the QD were in the
  stationary state at each instant of time. Also plotted is the spin component
  along the drain polarization $\vec{S} \cdot \hat{\vec{n}}_d$ (blue curves)
  computed from Eq. (\ref{eq:spr-kineqdiscuss}) for initial condition $\vec{S} =
  \hat{\vec{n}}_s / 2$ and $p_1 = 1 - p_0 = 1$. Throughout we used $n_s = n_d =
  0.99$ (see caption of Fig. \ref{fig:resonance}), $\alpha = 0.005 \pi$, $V_b
  = 50 T$, $W = 500 T$, $\Gamma_s = 0.15 T$, $\Gamma_d = 0.1 T$. The plots are
  obtained by numerically solving the analytically derived kinetic equations
  (\ref{eq:spr-kineqdiscuss}) in the limit $U \rightarrow \infty$ using the 
  scheme discussed in Appendix \ref{app:crossover}. To make use
  of analytical results, we need a tiny angle $\alpha$ here. For finite $U$,
  this restriction is unnecessary.\label{fig:timedep}}
\\
We compute the average current shown in Figs. \ref{fig:timedep}(c)--\ref{fig:timedep}(e) as follows: Taking the stationary state at $V_g^0$
as initial state $\rho (0)$, we obtain the time-dependent solution for $\rho
(t)$ by solving the kinetic equations (\ref{eq:spr-kineqdiscuss}). This yields
the time-dependent particle current $I_s (t)$ from Eq. (\ref{eq:spr-current}).
For both the current and the kinetic equations the rates are time-dependently
switched by changing the gate voltage $V_g^0 \leftrightarrow V_g$ in the
respective expressions according to the pulsing scheme. To ensure that Eq.
(\ref{eq:avcurr}) really gives the current measured in a circuit, we checked
that $\dot{p}_1 (t) \ll | I_s (t) |, | I_d (t) |$, i.e., the magnitudes of the
currents flowing out of the source, $| I_s (t) |$, and into the drain, $| I_d
(t) |$, are nearly the same. Under this condition displacement currents can be
neglected, as explained, for example, in
Ref.{\Cite{Schoeller97hab}}. We comment \new{in Appendix \ref{app:nonmarkov}} on the importance of
\New{non-Markovian corrections that we neglect here.}\\
The key feature of the current oscillations shown in Figs.
\ref{fig:timedep}(c)--\ref{fig:timedep}(e) is that the visibility strongly depends on the
voltage $V_g$ controlling the opening angle of the precession. \new{The
visibility} becomes
maximal at the resonance in Fig. \ref{fig:timedep}(d). \new{To prove our claim
that the current oscillations are correlated with a spin precession, we
compare in Figs. \ref{fig:timedep}(c)--\ref{fig:timedep}(e) the time-averaged current with
time-\emph{dependent} spin-projection curves, which are obtained as
follows: We take the
initial state $\rho (0)$ to be the maximally polarized state with spin
$\vec{S} = \op{\vec{n}}_s \text{/} 2$ and corresponding occupation
probabilities $p_1 = 1$ and $p_0 = 0$, i.e., we do not start from the stationary state at
gate voltage $V_g$. We then solve the kinetic equations
(\ref{eq:spr-kineqdiscuss}) time-dependently, keeping the gate voltage {\tmem{fixed}}
at $V_g$. The resulting spin vector $\vec{S}(t)$ is then projected on the drain polarization direction
$\op{\vec{n}}_d$, which yields the different spin projection curves
$\vec{S} (t) \cdot \hat{\vec{n}}_s$
in Figs. \ref{fig:timedep}(c)--\ref{fig:timedep}(e) with $\tau=t$. This comparison shows that the
current actually oscillates with the same frequency with which a spin would
precess in a QD held at gate voltage $V_g$.}\\
\new{Finally}, by going slightly off-resonance the precession {\tmem{axis}} can be fully tuned within
the plane of polarizations while maintaining full control over the precession
angle through $\tau$. This allows all single-spin operations required for
quantum algorithms to be implemented.\\
\subsection{Optimizing the pulsing scheme}\label{sec:optimal}
To set up an experiment that probes the underdamped spin precession, we
provide here some additional information under which conditions the
contrast of the signal obtained by the pulsing scheme is maximized. 
\new{For
this purpose, we discuss the ratio
$\bar{I} / \bar{I}_{\tmop{st}}$ of the time-averaged current
(\ref{eq:avcurr}) from the pulsing scheme, $\bar{I}$, to
the stationary current $\bar{I}_{\tmop{st}}$. The latter is obtained by
replacing the time-dependent current $I_s(t)=I_{\tmop{st}}(V_g(t))$ in
Eq. (\ref{eq:avcurr}) by the stationary current, switching only the gate
voltage $V_g$ as a parameter time-dependently.}
\\
First, underdamped precession cycles of a single spin are feasible only if the
spin-decay rate is much smaller than the spin-precession rate at the resonance
(see Sec. \ref{sec:shape}), that is, if
\begin{eqnarray}
  b \text{ \ } = \text{ \ } | B_{\perp} | \text{/} \sqrt{\Gamma_{\perp}
  \Gamma_{| |}} & \gg & 1. 
\end{eqnarray}
This condition is different from the condition that maximizes the stationary
current \New{, cf. Sec. \ref{sec:shape}}. There, we found a ratio $b \sim 1$ to be optimal because then roughly
one revolution takes place within the average electron dwell time on the QD.
If the tunneling rate allows for multiple precession cycles, the stationary
resonant current is suppressed again because tunneling happens infrequently,
even if its spin has optimal overlap with the drain polarization. This current
suppression near the resonance does not appear for the gate-pulsing scheme
since one returns to a gate voltage $V_g^0$ \New{closer to the single-electron tunneling resonance} where the tunneling rate is larger
and the electron can \New{leave} the QD quickly after it has been precessed
at gate voltage $V_g$. Thus, the larger the ratio $b$, the clearer
\New{the current oscillations are and the longer they persist.}\\
The second important set of parameters that has to be optimized are the dwell
times $\tau^0$ and $\tau$ at voltage $V_g^0$ and $V_g$. A first requirement is
that
\begin{eqnarray}
  \tau^0 & \lesssim & \tau \label{eq:tau0}
\end{eqnarray}
because if $\tau^0 \gg \tau$ the system is most of the time not at resonance
and the average current is determined by the dynamics at gate voltage
$V_g^0$.
Condition (\ref{eq:tau0}) is fulfilled for most values of $\tau$ shown in Fig. \ref{fig:timedep}.
However, there is another condition that is equally important: We find on a
numerical basis that $\tau^0$ is chosen optimally as
\begin{eqnarray}
  \tau^0 & \approx & 0.1 \tau_T^0 \text{ \ } \approx \text{ \ } 0.1 /
  I^0_{\tmop{st}}, 
\end{eqnarray}
with the electron dwell time $\tau_T^0$ at gate voltage $V_g^0$, which can be
estimated by the inverse of the stationary particle current $I^0_{\tmop{st}}$
at voltage $V_g^0$. If $\tau^0  \gtrsim \tau_T^0$, the average current is
mostly determined by the large {\tmem{stationary}} current $I^0_{\tmop{st}}$,
i.e., the precession-induced initial modification of the current at $V_g^0$ is
rather insignificant. This is illustrated in Fig. \ref{fig:prectauTwo}(a), in which we plot the ratio of the average current
$\bar{I}$ obtained from the pulsing scheme and the average current
$\bar{I}_{\tmop{st}}$ that is obtained if the QD was in the stationary state all the
time (but switching between the different levels at the two gate voltages):
Clearly, for small (but not very small) times $\tau^0$, the current is
drastically enhanced over the stationary current due to the gate pulsing,
while the ratio decreases if $\tau^0$ approaches $\tau_T^0$.
In Fig. \ref{fig:timedep}, we use a value $\tau^0/\tau^0_T \sim 1$, which already yields a sizable enhancement.
  \Figure{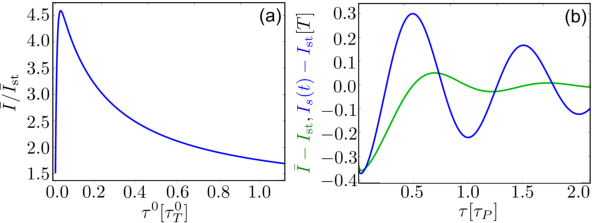}{(a) Ratio $\bar{I} \text{/} \ \bar{I}_{\tmop{st}}$ as a
  function of the duration $\tau^0$ in units of \ the electron dwell time
  $\tau_T^0 = 1 / I_{\tmop{st}}^0 \approx 4.7 \cdot 10^4 / T$. Here, $\bar{I}$
  is the average current (\ref{eq:avcurr}) for the pulsing scheme and
  $\bar{I}_{\tmop{st}}$ is the stationary current obtained if the QD was in the
  stationary state all the time. The gate voltage $V_g = V_g^{\ast} = 59.8 T$
  is tuned to the resonance and $\tau = 2500 \text{/} T \approx 0.53 \tau_P$
  so that a nearly maximal enhancement of the current occurs after the
  precession. (b) Average current $\bar{I}$ and time-dependent
  current $I_s (t)$ as function of the pulse time $\tau$ with
  $\tau_P \approx 4.7 \cdot 10^3 / T$ and $\tau^0 = 10 \text{/} T = 2.1 \cdot
  10^{- 3} \tau_P \ll \tau$. We subtract the stationary current
  $I_{\tmop{st}}$ flowing at gate voltage $V_g=V_g^{\ast}$. The results shown in (a) and (b) are computed for
  a single pulse, $N = 1$ [$t = \tau^0 + \tau$ in Eq. (\ref{eq:avcurr})], and
  all other parameters are the same as in Fig. \ref{fig:timedep}.\label{fig:prectauTwo}}
\\
However, if $\tau^0 \ll \tau_T^0$, the ratio $\bar{I} /
I_{\tmop{st}}$ becomes drastically suppressed as Fig. \ref{fig:prectauTwo}(a) also shows. In this case, the QD electron does not have
enough time to tunnel out of the QD when the gate voltage is switched to
$V_g^0$. The average current is then mostly determined by the time-averaged
current at resonance. This is illustrated in Fig. \ref{fig:prectauTwo}(b), which shows the time-dependent current $I_s (t)$
(blue) besides its average current $\bar{I}$(green), which loses contrast
after roughly two cycles.\New{
\\
We} conclude that for setting up and optimizing the pulsing scheme in
an experiment, the initial characterization of the spin valve is of the utmost
importance. Once the time scales are known our above discussion should be a
guide to choose the pulsing times properly.\\
\subsection{Experimental feasibility: spin decoherence}\label{sec:estimate}
\New{Finally, we provide rough} estimates for the spin-decay times and spin-precession periods
for experimentally achievable parameters, demonstrating the feasibility of
underdamped spin precession cycles in the Coulomb blockade regime. Typical
spin-dephasing times of $\sim 10 - 30 \tmop{ns}$ have been measured in GaAs
QDs  \cite{Petta05,Koppens08,Bluhm10} and are also compatible with
measurements involving carbon nanotubes (CNTs) \cite{Hueso07}. In our case,
the cotunneling current through the QD leads to additional dephasing with time
constant $\sim U / \Gamma^2 \sim 10 / \mu\tmop{eV} \sim 30 \text{ns}$ for
typical values of $\Gamma \sim 0.01 \tmop{meV}$ and $U \sim 5 \tmop{meV}$
feasible both for semiconductor QDs and CNT QDs. The energy scale related to
the exchange field may be estimated as $\mu_B B > \mu_B B_{d, \perp} \approx
\mu_B | \log (1 / 2) | \Gamma_d n_d \sin \alpha / \pi \sim 0.7 \mu \tmop{eV}$
for $n_d \sim 0.5$ and $\alpha \sim 0.2 \pi$. This translates into a maximal
\New{precession} period of $T \sim 2 \pi / 0.7 \mu \tmop{eV} \sim 6 \tmop{ns}$ at the resonance
and even smaller periods away from it. Thus, indeed, the spin precession
period can be made smaller than the spin-decay time. \\
\new{One may wonder whether the spin resonance could also be observed in
the strong-coupling regime $\Gamma \gg T$. This regime has been under intense experimental
investigation (using collinear polarizations so far) since the exchange
field can be probed there by the strong
spin splitting it induces, affecting the Kondo resonance \cite{Hamaya07,Hauptmann08,Gaass11}.
Increasing the tunnel coupling $\Gamma$, however, enhances the
spin-decoherence rate more strongly than the spin-precession rate. Moreover, spin-flip processes driving the Kondo
effect for small bias voltage also destroy the QD spin coherence. 
Thus, the spin precession may not be underdamped any more in the
strong coupling regime. In addition, the smaller spin-precession
periods, which are approximately $\sim$100 ps as extracted from
exchange-field magnitudes in Ref.{\Cite{Hauptmann08}}, makes it more challenging
to apply the pulsing scheme described above. By contrast, spin-resonance
features are more likely to be seen
in the \emph{stationary} current, which requires the spin-precession rate
only to be comparable to the spin-decoherence rate. The reader should
note that the width of the spin resonance can be much smaller than
the spin-decoherence rate as our analysis in Sec. \ref{sec:shape} shows.
Thus, it is worth investigating the spin resonance in the
strong-coupling regime further. We finally note that in the meantime of revising the manuscript of this paper, signatures of the spin resonance have also been found by one of the authors in waiting-time distributions \cite{Sothmann14}.}
\\
\section{Summary and outlook}\label{sec:spr-summary}
We have identified a spin resonance that, unlike usual resonances, does not
appear when scalar energies of the local quantum system and reservoir match.
Instead, the condition (\ref{eq:resonance}) based on vectors, $\vec{B} \cdot
\hat{\vec{n}}_s = 0$, needs to be satisfied. The resonance emerges in the simplest
QD spin-valve setup one can think of: an interacting spin-degenerate single
level which is tunnel-coupled to two ferromagnets for {\tmem{almost}} (but not
exactly) antiparallel \new{polarizations} $\vec{n}_s$ and $\vec{n}_d$. For this
magnetic configuration, the direction of the exchange field $\vec{B}$
strongly depends on the applied voltages, which generates a sharp feature
\New{all across} the Coulomb diamond of the transport stability diagram.\\
\newer{The resonance is clearly distinguished from other features in the stability
diagram: First, it emerges only
for nonzero noncollinearity angle $\alpha$ and responds sensitively to  
changes in $\alpha$. Second, it depends strongly on the  asymmetries of the spin-injection
rates.
For small asymmetries, it exhibits a strong
current rectification effect \New{while for symmetric
\New{spin-injection} rates it lies at zero bias.}
Third, when these parameters cannot be controlled in an experiment, one can use the peculiar voltage-dependent line shape of the spin resonance to tell it apart
from other features. For example, the spin resonance vanishes at the particle-hole symmetry point.
Furthermore, its width is not given by a simple combination of the tunnel couplings,
temperature, or any other natural energy scale. Instead, it depends on
the ratio of the gate-voltage dependent exchange field and the spin-decay
rates. The latter features contrast particularly with those from the Kondo effect or the zero-bias anomaly predicted earlier in
Refs.{\Cite{Weymann05} and\Cite{Weymann05b}}   }.\\
While the resonance position is entirely dictated by the exchange field
{\tmem{direction}}, the shape of the resonance (peak value, width) is strongly
influenced by the QD spin decay. We have identified the ratio $b = | B_{\perp}
| / \sqrt{\Gamma_{\perp} \Gamma_{| |}}$ to be the relevant parameter that
determines the resonance shape. \New{This ratio} involves the perpendicular
exchange field component $B_{\perp} = \vec{B} \cdot \op{\vec{n}}_{\perp}$, the
spin-relaxation rate $\Gamma_{| |} = \op{\vec{n}}_s \cdot \mathcal{R}_S \cdot
\op{\vec{n}}_s$, and the spin-dephasing rate $\Gamma_{\perp} =
\op{\vec{n}}_{\perp} \cdot \mathcal{R}_S \cdot \op{\vec{n}}_{\perp}$. The
resonance appears for $b \gtrsim 0.1$ in the stationary transport, which is
satisfied in the Coulomb-blockade regime. There, the spin decay is limited by
next-to-leading order processes \New{(cotunneling)} with rate $\Gamma_{\perp, | |}
\sim \Gamma^2 / U$, which can indeed be made smaller \new{by reducing
the tunnel couplings} than the spin-precession
frequency $B_{\perp} \sim \Gamma$ at resonance. The strongest contrast in the
stability diagram is expected for $b \approx 1$ when roughly half a spin
revolution happens within the electron dwell time in the QD. By contrast, in
the limit $b \gg 1$, the spin coherence lasts much longer than one spin
revolution and underdamped spin-precession cycles becomes feasible.
\New{The underdamped spin precession} leads
to no qualitative modifications of the resonance in the stationary
conductance, but \New{can nevertheless be probed}
experimentally by a simple gate-pulsing scheme. In the latter, the precession
axis is controlled by electrical means and the rotation angle by the duration
of the pulses. This allows one to determine the magnitude of the exchange
field and in combination with stationary-conductance measurements to determine
the \new{anisotropy of the spin decay all-electrically}. This can even be used to
realize \New{every} single-spin qubit \New{gate operation} in an all-electric way.\\
Besides opening new avenues for spintronics and single-spin control, the spin
resonance studied \New{here provides} an illustration of a generic concept in
the simplest conceivable setting: such an anomalous resonance can appear in
any open \New{quantum} system with quasi-degenerate states whose coherence is described by a
Bloch vector. For a two-level system, it is required that (i) the Bloch vector
suffers only from little decoherence, (ii) the coherent evolution is dominated
by a renormalization-induced field vector -- because of level degeneracy --
(iii) which is induced by an environment that breaks symmetries (which are
often present in idealized models). \New{When such a system is tuned by experimentally}
accessible parameters, \New{a} resonance unrelated to any energy
splitting \New{can appear} when the field vector \New{becomes} perpendicular to the Bloch vector.\\
This can be extended to $N$-fold degenerate multiplets, described by a
generalized Bloch vector and an associated renormalization field vector.
Interestingly, in this case the precession takes place in a higher dimensional
space and is expected to be overlooked even more easily as compared to the
simple case studied here. Scenarios can be envisaged in nuclear spin systems,
\cite{Slichterbook} double QDs \cite{Wunsch05}, or vibrating molecular
devices \cite{Donarini06,Schultz10,Koller10}. Our simple example shows
\New{that, interestingly, Coulomb} interactions realize both requirements (i) and (ii) while noncollinear spin
valves naturally provide (iii).\\
\section*{Acknowledgements}
We acknowledge discussions with M. Baumg{\"a}rtel, S. Das,
S. Herzog, M. Misiorny,
and F. Reckermann,
and financial support of the Swiss National Science Foundation via the NCCR QSIT [B.~S.], and the Swedish Research
Council (VR) [M.~L.].
\begin{appendix}
 
\section{Quantum master equations and Pauli
superbasis}\label{app:liouville}
\New{In this Appendix, we \New{outline} how our coupled differential equations for operator averages
(\ref{eq:spr-kineqdiscuss}) can be obtained from the general kinetic equation (\ref{eq:kineq}) 
for the reduced operator of the QD.} For this purpose, we use a Liouville-space notation whose key elements we briefly
review.\\
\new{Applying} the real-time diagrammatic technique \cite{Leijnse08a,Schoeller09a},
one can express the kinetic equation of the reduced density operator $\rho
(t)$ of the QD as
\begin{eqnarray}
  | \dot{\rho} (t)) & = & - i L| \rho (t)) + \int_{- \infty}^t d t' W (t - t')
  | \rho (t')).  \label{eq:kineqgeneral}
\end{eqnarray}
Here, we introduced a bra-ket notation for linear operators $|A) :
\mathcal{H} \rightarrow \mathcal{H}$ acting on a Hilbert space $\mathcal{H}$, which
altogether form the Liouville space $\mathcal{L}$. Furthermore, $L$
and $W$ denote {\tmem{super}}operators, which are operators
\new{$\mathcal{S} : \mathcal{L} \rightarrow \mathcal{L}$} mapping a
Liouville-space element on another Liouville-space element. In particular, $L\bullet = [H, \bullet]$ mediates the internal evolution of the \New{density}
operator by the QD Hamiltonian $H$, where the dot ``$\bullet$'' denotes the
operator that $L$ acts on. Furthermore, $W (t - t')$ is the real-time
diagrammatic kernel that incorporates the effect of the environment on the
evolution of the reduced system. We next carry out a Markov approximation,
that is, we replace $| \rho) (t') \approx | \rho) (t)$ in Eq.
(\ref{eq:kineqgeneral}) and obtain
\begin{eqnarray}
  | \dot{\rho} (t)) & = & [- i L + W] | \rho (t)) .  \label{eq:markov}
\end{eqnarray}
Here, $W = \int_0^{\infty} d t e^{i z t} W (t) |_{z = i 0}$ is the
zero-frequency component of the kernel. One can prove
\cite{Schoeller09a,Leijnse08a} that the stationary state calculated from Eq.
(\ref{eq:markov}) is the {\tmem{exact}} stationary solution of Eq.
(\ref{eq:kineqgeneral}).
For actual calculations, Eqs. (\ref{eq:kineqgeneral}) and (\ref{eq:markov})
are expressed in terms of matrix elements. To achieve this goal, one
introduces the following scalar product in Liouville space: \cite{Breuer}
\begin{eqnarray}
  (A  |B) & := & \tmop{Tr} (A^{\dag} B) . 
\end{eqnarray}
An orthonormal {\tmem{superbasis}} is a set of superstates $\{ |A) \}$ that is
orthonormal with respect to this scalar product,
\begin{eqnarray}
  (A  |B) & = & \delta_{A B}, 
\end{eqnarray}
and satisfies the completeness relation:
\begin{eqnarray}
  \mathcal{I} & = & \sum_A |A) (A|.  \label{eq:completeness}
\end{eqnarray}
Here, $\mathcal{I}$ denotes the superidentity $\mathcal{I} | A) = | A)$ for
any $|A)$. As a consequence, any superstate $| O)$ can be expanded into such
an orthonormal basis by $| O) = \sum_A^{} O_A | A)$ with coefficients $O_A =
(A  |O)$ and any superoperator can be expressed as
$\mathcal{S} = \sum^{}_{A, B} S_{A B} | A) (B |$ with
$\mathcal{S}_{A B} = (A| [\mathcal{S} |B)]$. \\
Usually, Eq. (\ref{eq:markov}) is expressed in terms of matrix elements for
the superbasis $|a, b) : = { |} a { \rangle}
{ \langle} b { |}$, which yields
\begin{eqnarray}
  \dot{\rho}_{a b} &  & - i L^{a b}_{a' b'} \rho_{a' b'} + W^{a b}_{a' b'}
  \rho_{a' b'},  \label{eq:kineqmatrix}
\end{eqnarray}
with $\rho_{a b} = (a, b|  \rho) = \tmop{Tr} ([{ |} a
{ \rangle} { \langle} b {|}]^{\dag} \rho)$ and $\mathcal{S}^{a b}_{a' b'} = (a, b|
\mathcal{S} |a', b') = \tmop{Tr} ([{ |} a
{ \rangle} { \langle} b {
|}]^{\dag} [\mathcal{S} { |} a'
{ \rangle} { \langle} b' { |}])$ for $\mathcal{S} = L, W$. In the Keldysh-contour
formulation of real-time diagrammatics,  \cite{Koenig96b,Koenig96prl} diagram
rules are given for the kernel matrix elements $W^{a b}_{a' b'}$. The diagonal
matrix elements $\rho_{a a}$ are interpreted as occupation probabilities and
the off-diagonal elements $\rho_{a b}$ as coherences. For a different choice
of the basis states, however, the coherences in the former basis contribute to
the occupation probabilities in the new basis. Thus, the interpretation as
``probabilities'' and ``coherences'' is meaningful only if a specific basis is
singled out by the symmetry of the problem.\
For the single-level Anderson model we consider here, this would be the
case for nonmagnetic electrodes. \New{In this case,} one can start from the Hilbert space
basis \{${ |} 0 { \rangle}$, ${|} \uparrow { \rangle}$, ${ |} \downarrow
{ \rangle}$, ${ |} 2 {\rangle}$\} with a fixed quantization axis for the spin. All coherences
between spin states are zero in the stationary limit.\\
For noncollinear lead polarizations, such a spin quantization axis does not \New{exist}, c.f.,
Sec. \ref{sec:sprmodel}. Thus, it is helpful to expand Eq. (\ref{eq:markov}) in terms of different supermatrix elements such that all
expressions \New{are independent of the quantization axis.}
For this purpose, we chose a superbasis $\{ |A) \}$ consisting of
{\tmem{observables}}. The reduced density matrix can then be expanded as
\begin{eqnarray}
  | \rho) & = & \sum_A A |A),  \label{eq:densexp}
\end{eqnarray}
where
\begin{eqnarray}
  A & = & (A  | \rho) = \tmop{Tr} (A \rho) 
\end{eqnarray}
is the expectation value of observable $A (= A^{\dag})$ - an object with an intuitive
\New{physical} interpretation in contrast to the matrix elements
$\rho_{a b}$.\\ 
For the single-level Anderson model, a suitable superbasis is the {\tmem{Pauli
superbasis}}. We focus here on the charge-conserving setup without
superconductor (see Fig. \ref{fig:model}). For the
nondegenerate subspaces with zero ($n = 0$) and two $(n = 2)$ electrons, these
are simply the projectors
\begin{eqnarray}
  | \check{r}^0_0) & := & \op{P}_0 \text{ \ } = \text{ \ } | 0 \rangle
  \langle 0 |
\end{eqnarray}
and
\begin{eqnarray}
  | \check{r}^2_0) & := & \op{P}_2 \text{ \ } = \text{ \ } | 2 \rangle
  \langle 2 | .
\end{eqnarray}
For the subspace of charge state $n = 1$, we introduce
\begin{eqnarray}
  | \check{r}_{\mu}^1 ) &  : = & \sum_{\sigma \sigma'}
  (\check{r}_{\mu})_{\sigma \sigma'} | \sigma \rangle \langle \sigma' |,
  \label{eq:toperators}
\end{eqnarray}
where, $\mu = 0, 1, 2, 3$, $(\check{r}_0)_{\sigma \sigma'} = \delta_{\sigma
\sigma'} \text{/} \sqrt{2}$, and \ $(\check{r}_{\mu = i})_{\sigma \sigma'} =
(\sigma_i)_{\sigma \sigma'} \text{/} \sqrt{2}$ involving the Pauli matrices
$\sigma_i$ for $i = 1, 2, 3$. The element $| \check{r}_0^1 ) =
\op{P}_1 / \sqrt{2}$ is proportional to the scalar projection operator on
charge state 1 and the elements $| \check{r}_i^1 ) = \sqrt{2}
\op{S}_i$ are proportional to the vector components of the spin operator.
Altogether, these six superstates provide an orthonormal basis for the
subspace of the charge-diagonal QD operators,
\begin{eqnarray}
  (\check{r}^n_{\mu}  | \check{r}_{\mu'}^{n'}) & = & \delta^{n n'}
  \delta_{\mu \mu'},  \label{eq:orthonormal}\\
  \mathcal{I}_C & = & \sum_{n \mu} | \check{r}_{\mu}^n )
  (\check{r}^n_{\mu} |,  \label{eq:identity}
\end{eqnarray}
where $\mathcal{I}_C$ denotes the identity operator in the subspace of
charge-diagonal operators. The factors $1 \text{/} \sqrt{2}$ are introduced in
the definition of $(\check{r}_{\mu})_{\sigma \sigma'}$ to avoid additional
factors in Eqs. (\ref{eq:orthonormal}) and (\ref{eq:identity}).\
The Pauli superbasis is sufficient to expand the density operator $|
\rho)$ \footnote{In principle, ten more superbasis elements have to be added to form a complete
set of all superstates of the Liouville space of the QD, which has dimension
$4^2 = 16$, because of the four-dimensional Hilbert space considered here.
However, for expanding any observable or the density operator, eight of these
superbasis elements are obsolete since they do not conserve the fermion parity
\cite{Saptsov14}. Furthermore, we restrict our considerations here to the
charge-conserving model without a superconductor.} which
reads by applying Eq. (\ref{eq:densexp}):
\begin{eqnarray}
  | \rho) & = & \tfrac{1}{\sqrt{2}}  \sum_n p_n | \check{r}^n_0) + \sqrt{2} \
  \vec{S} \cdot | \check{\vec{r}}^1),  \label{eq:expand}
\end{eqnarray}
where $p_1 = \sqrt{2} \tmop{Tr} (\check{r}^1_0 \rho)$, $p_{0 \text{/} 2} =
\tmop{Tr} ( \check{r}^{0 \text{/} 2}_0 \rho )$ are the
occupation probabilities of charge state $n$ and $\vec{S} = \tmop{Tr} \left(
\check{\vec{r}}^1 \rho \right) \text{/} \sqrt{2}$ is the average spin operator
(\ref{eq:sd}). Importantly, Eq. (\ref{eq:expand}) is {\tmem{covariant}}, i.e.,
form-invariant under a change of the spin-quantization axis or the real-space
coordinate system. This also illustrates that working in Liouville space does
not only give more compact expressions but it also results in a physically more
transparent description of the QD state and its dynamics.
\section{Kinetic equations and current for quantum-dot spin
valve}\label{app:kineq}
\subsection{Rates for kinetic equations and current}\label{app:rates}
In this Appendix, we give all expressions for the rates in the kinetic
equations (\ref{eq:spr-kineqdiscuss}), which read:
\begin{eqnarray}
  & \begin{array}{lll}
    \dot{p}_0 & = & - 2 \Gamma_0 p_0 + \Gamma_1 p_1 + 2 \vec{G}_{p S} \cdot
    \vec{S},\\
    \dot{\vec{S}} & = & + \vec{G}^0_{S p} p_0 - \tfrac{1}{2} \vec{G}^1_{S
    p} p_1 - \mathcal{R}_S \cdot \vec{S} - \vec{B} \times
    \vec{S} .
  \end{array} &  \label{eq:app-kineq}
\end{eqnarray}
The charge-relaxation rates are given by
\begin{eqnarray}
  \Gamma^{0 \text{/} 1} & = & \Gamma^{\pm}_0 \pm \tmop{Im} (K^+_{0 0}
   + \tfrac{1}{2} \sum_{\rho}  K^-_{\rho \rho}), 
  \label{eq:cot}
\end{eqnarray}
where Greek indices take the values $\rho = 0, 1, 2, 3$ and Latin indices take
the values $i = 1, 2, 3$. Furthermore, the vectorial spin-to-charge conversion
rates are given by
\begin{eqnarray}
  (G_{p S})_i & = & \Gamma_i^- - \tmop{Im} \left. \left( K^+_{i 0} +
  \tfrac{1}{2} K^-_{i 0} + \tfrac{1}{2} K^-_{0 i} \right) \right. \nonumber\\
  &  & - \tfrac{1}{2} \sum_{j k} \varepsilon_{i j k} \tmop{Re} (K^-_{j k}), 
  \label{eq:gps}
\end{eqnarray}
the vectorial charge-to-spin conversion rates are given by
\begin{eqnarray}
  (G^{0 \text{/} 1}_{S p})_i & = & \Gamma_i^{\pm} \pm \tmop{Im} \left.
  \left( K^+_{i 0} + \tfrac{1}{2} K^-_{i 0} + \tfrac{1}{2} K^-_{0 i} \right)
  \right. \nonumber\\
  &  & \mp \tfrac{1}{2} \sum_{j k} \varepsilon_{i j k} \tmop{Re} (K^-_{j k}),
  \label{eq:gsp}
\end{eqnarray}
the symmetric spin-decay tensor is defined by
\begin{eqnarray}
  (\mathcal{R}_S)_{i j} & = & \delta_{i j} \Gamma^-_0 +
  \delta_{i j} \tmop{Im} \left( - \tfrac{1}{2} K^-_{0 0} + \tfrac{1}{2} \sum_i
  K^-_{i i} - D^{- +}_{0 0} \right) \nonumber\\
  &  & - \tfrac{1}{2} \tmop{Im} (K^-_{i j} + K^-_{j i} + X^{+ -}_{i
  j} + X^{+ -}_{j i}),  \label{eq:rsij}
\end{eqnarray}
and, finally, the vectorial exchange field reads
\begin{eqnarray}
  B_i & = & \beta_i + \tmop{Re} \left. \left( \tfrac{1}{2} K^-_{i 0} -
  \tfrac{1}{2} K^-_{0 i} + D^{- +}_{i 0} \right) \right. .  \label{eq:fullb}
\end{eqnarray}
The above rates first contain terms of $O (\Gamma)$, $\Gamma_{\rho}^{\chi}
(\varepsilon) = \sum_r \Gamma_{r, \rho}^{\chi} (\varepsilon)$ and
$\beta_{\rho} (\varepsilon) = \sum_r \beta_{r,\rho} (\varepsilon)$, with
\begin{eqnarray}
  \Gamma_{r, \rho = 0}^{\chi} (\varepsilon) & = & \Gamma_r^{\chi}
  (\varepsilon) = 2 \pi | t_r |^2 \bar{\nu}^r f (\chi \text{} (\varepsilon -
  \mu_r) \text{/} T),  \label{eq:gamma}\\
  \Gamma^{\chi}_{r,\rho = i} (\varepsilon) & = & \Gamma^{\chi}_r
  (\varepsilon) n_{r,i} \text{ \ } (i = 1, 2, 3), \\
  \beta_{r,\rho} (\varepsilon) & = & P \int_{- W}^{+ W} \frac{d
  \omega}{\pi}  \frac{\Gamma_{r,\rho}^{+} (\omega)}{\varepsilon - \omega}, 
  \label{eq:bfield}
\end{eqnarray}
with $P$ denoting the principal value integral and the Fermi function $f (x) =
1 \text{/} (e^x + 1)$. Here, the spatial components $\rho = 1, 2, 3$ point
along by the polarization vector $\vec{n}_r$ of lead $r$.
Furthermore, the $O (\Gamma^2)$ contributions incorporate two different
tensors, namely
\begin{eqnarray}
  & & X^{\chi_2 \chi_1}_{\rho_2 \rho_1} = \int_{- W}^{+ W}  \int_{- W}^{+ W}
  \frac{d \omega_1}{\pi} \frac{d \omega_2}{\pi}
  \Gamma_{\rho_2}^{\chi_2} (\omega_2) \Gamma_{\rho_1}^{\chi_1} (\omega_1)
  \nonumber\\
  &  & \text{ \ \ \ \ \ \ \ }\frac{1}{i 0 + \omega_2 - \varepsilon}  \frac{1}{i 0 + \omega_2 -
  \omega_1} \frac{1}{i 0 - \omega_1 + \varepsilon},  \label{eq:xint}
\end{eqnarray}
and $D^{\chi_2 \chi_1}_{\rho_2 \rho_1}$ given by the same expression when
replacing the right-most denominator in the \new{above} expression by $\text{1} \text{/}
[i 0 + \omega_2 - \varepsilon]$.
\New{In contrast to previous works, we evaluate the full complex integral
to completely capture the dynamics of the \new{spin} coherences in the Coulomb blockade regime to order $\Gamma^2$.}
Adding the $X$- and $D$-integrals, we obtain
the \New{simpler} function
\begin{eqnarray}
  K_{\rho_2 \rho_1}^{\chi_1} & = & \bar{\chi}_2 (X^{\chi_2 \chi_1}_{\rho_2
  \rho_1} + D^{\chi_2 \chi_1}_{\rho_2 \rho_1}) \\
  & = & [\chi_1 \beta'_{\rho_2} \beta_{\rho_1} + \Gamma_{\rho_2}'
  \Gamma^{\chi_1}_{\rho_1}] + i [\chi_1 \Gamma_{\rho_2}' \beta_{\rho_1} -
  \beta'_{\rho_2} \Gamma^{\chi_1}_{\rho_1}],  \nonumber \label{eq:kdec}
\end{eqnarray}
where $\Gamma'_{\rho} = d \Gamma^+_{\rho} \text{/} d \varepsilon$
and $\beta'_{\rho} = d \beta_{\rho} \text{/} d \varepsilon$ .\\
\newer{We note that the leading-order $\Gamma$ contribution to
the spin-relaxation tensor (\ref{eq:rsij}) is isotropic while the next-to-leading order $\Gamma^2$
contribution renders the spin decay anisotropic.
Since the leading-order term is suppressed
in the Coulomb-blockade regime, the spin decay can indeed become significantly
anisotropic. \new{In contrast to the decay rates,} the first-order $\Gamma$
contribution to the exchange field
(\ref{eq:bfield}) is only logarithmically suppressed.}\\
\New{The} expression for the average current from lead $r$ into the QD
reads
\begin{eqnarray}
  I_r & = & 2 \Gamma_{r,0} p_0 - \Gamma_{r,1} p_1 - 2 \vec{G}_{r,p S} \cdot
  \vec{S}  \label{eq:spr-current-app}
\end{eqnarray}
with
\begin{eqnarray}
  \Gamma_{r,0 / 1} & = & \Gamma^{\pm}_r \pm \tmop{Im} (K^{+}_{r,0 0})
  + \tfrac{1}{2}  \sum_{\rho} \tmop{Im} (K^{-}_{r, \rho \rho}), \\
  (G_{r,p S})_i & = & \Gamma_{r,i}^{-} - \tmop{Im} \left. \left( K^{+}_{r,i 0}
  + \tfrac{1}{2} K^{-}_{r,i 0} + \tfrac{1}{2} K^{-}_{r,0 i} \right) \right.
  \nonumber\\
  &  & + \tmop{Im} (X^{+ -}_{r,0 i} - X^{- +}_{r,i 0})
  \nonumber\\
  &  & - \tfrac{1}{2}  \sum_{\rho_2 \rho_1} \varepsilon_{i \rho_2 \rho_1}
  \tmop{Re} (K^{-}_{r,\rho_2 \rho_1}),  \label{eq:gpsr}
\end{eqnarray}
where $X^{\chi_2 \chi_1}_{r,\rho_2 \rho_1}$ is obtained from Eq.
(\ref{eq:xint}) by replacing $\Gamma_{\rho_2}^{\chi_2} (\omega_2) \rightarrow
\Gamma_{r, \rho_2}^{\chi_2} (\omega_2)$ and $K^{\chi}_{r, \rho_2 \rho_1}$ is
obtained from Eq. (\ref{eq:kdec}) by replacing $\beta'_{\rho_2} \rightarrow
(\beta_{r,\rho_2})'$ and $\Gamma'_{\rho_2} \rightarrow (\Gamma_{r,\rho_2})'$,
respectively.\\
\newer{The $X$-type integrals, Eq. (\ref{eq:xint}), and the \New{corresponding} $D$-type
integrals are computed numerically as we explain next. We convert the double frequency integral into a double summation over Matsubara frequencies by first substituting $x_2$=$\omega_2/T$ and $x_1$=$-\omega_2/T$ and splitting the Fermi functions $f(xT)=g^+(x) + g^-(x)$ into their symmetric part $g^+(x)=1/2$ and their antisymmetric part  $g^-(x)=-\tanh(x/2)/2$, respectively. We then integrate over $x_1$ and $x_2$ using complex integration, closing the integration contour in the upper half of the complex plane. By virtue of the residue theorem, one can derive the following relation for the generic type of integrals occurring after these manipulations
    \begin{eqnarray}
    & & \int_{-R}^{+R}\!\! \textrm{d}x_1\!
    \int_{-R}^{+R}\!\!  \textrm{d}x_2 \
    g^{q_1}(x_1) g^{q_2}(x_2) \nonumber \\
    & &
    \frac{1}{x_j-\lambda_2+i0}
    \frac{1}{x_1+x_2+i0}
    \frac{1}{x_1-\lambda_1+i0} \nonumber \\
    & =\ &- 4 \pi^2\delta_{q_1,-}\delta_{q_2,-}
    \sum_{k_1,k_2}^{k_R}
   \frac{1}{z_{k_j} - \lambda_2 }
   \frac{1}{z_{k_1} + z_{k_2} }
    \frac{1}{z_{k_1} - \lambda_1 } \ \ \ \nonumber \\
   & & + 2 \pi i \delta_{q_1,-} \delta_{j,1} \sum_{k_1}^{k_R}
   \frac{1}{z_{k_1} - \lambda_2 }\frac{1}{z_{k_1} - \lambda_1 }
   \sum_{k_2}^{k_R} M^{q_2}_{k_2} \nonumber
      \\
   & & + O\left(\frac{1}{R}\right),
    \label{eq:int}
    \end{eqnarray}
where $j=1,2$, $q_1,q_2 = \pm$, $z_{k_{1,2}} = i \pi (2k_{1,2}+1)$ are the Matsubara frequencies, and $0 \leq k \leq k_R = \lceil \tfrac{R}{2 \pi} - \tfrac{1}{2} \rceil$ with $\lceil x \rceil$ denoting the smallest integer that is not less than $x$. We additionally used the abbreviation
    \begin{align}
    M^{q_2}_{k_2} =
    \frac{1}{2} \left[ q_2 \ln\left( \frac{z_{k_2}+i R }{z_{k_2} + R}  \right) + \ln\left( \frac{z_{k_2}-R }{z_{k_2} + i R}  \right) \right].
    \end{align}
The above double Matsubara sums are then evaluated numerically.}
\subsection{Extension of former studies}\label{app:extension}
In this \New{Appendix}, we compare our kinetic equations
(\ref{eq:spr-kineqdiscuss}) to \New{those of} prior studies of QD
spin valves \New{and results from other approaches}: In fact, our theoretical approach \New{presents a
technical step forward relative to the} previous works, which is a reason why the spin
resonance has been overlooked for a long time.\\
\New{\emph{Quamtum master equations.}}
First, the lowest-order $\Gamma$ contribution to our equations
complies with the results given in Refs.{\Cite{Koenig03} and\Cite{Braun04set}} taking
the limit $U \rightarrow \infty$. However, a lowest-order expansion in
$\Gamma$ is not sufficient in the Coulomb blockade regime since these terms
are exponentially suppressed with the distance $| \varepsilon - \mu_r | / T$
from the Fermi levels. By contrast, \New{some} $O (\Gamma^2)$-terms are only
algebraically suppressed and therefore dominate there. \New{In particular, this} is associated with
a spin decay \New{due to cotunneling} that could obliterate the coherent spin-precession
features. \New{This was noted in Ref.{\Cite{Baumgaertel11}} where the spin resonance
was reported to emerge on the flank of the Coulomb diamond using an $O (\Gamma)$ kinetic
equation, but could not be reliably followed into the Coulomb blockade regime.}
However, the sharp resonance feature we find here even when $O
(\Gamma^2)$ cotunneling corrections are included shows that spin precession
effects can still be dominant -- as anticipated in the introduction from
time-scale estimations.\\
Next-to-leading order corrections $\sim \Gamma^2$ have been included \New{in other studies of the same model,}
 for example, in Refs.{\Cite{Weymann05}, \Cite{Weymann05b}, and \Cite{Koller12}}; however
these works address only \new{\emph{collinearly}} polarized ferromagnets. Here, the spin
precession cannot occur since the spin accumulation and the exchange
field are collinear (cf. the expressions of the rates in
Appendix \ref{app:rates}). In Ref.{\Cite{Weymann05c}}, also the
noncollinear magnetic configuration is studied, but the QD is assumed to be
deposited on a ferromagnetic substrate causing a large splitting
$\varepsilon_{\uparrow} - \varepsilon_{\downarrow} \gg \Gamma$ of the two
spin states as compared to the tunnel coupling, so that the spin components
transverse to this splitting field have negligible impact.\\
The \New{difficult} case of degenerate QD spin states, noncollinear polarizations, and
cotunneling corrections has to our knowledge been addressed only in
Ref.{\Cite{Weymann07c}}. While the kinetic equations given
there include all the terms that correspond to the imaginary parts up to $O
(\Gamma^2)$ and the real parts up to $O (\Gamma)$ \newer{in the rates
(\ref{eq:cot}) - (\ref{eq:fullb})}, our
equations additionally include the $O (\Gamma^2)$ corrections to the real
parts \newer{of these rates. This is done \New{via} the Matsubara double summation
(\ref{eq:int}), which is implemented numerically. For other \New{models with higher degree of symmetry}, which
only require the imaginary part of these integrals, this can be avoided
(see Ref.\Cite{Leijnse08a}).
Thus, we include, for example, in the exchange field (\ref{eq:fullb})} {\tmem{all renormalization effects}} up to $O(\Gamma^2)$. Our results actually confirm that the $O (\Gamma^2)$ corrections to the exchange field are \New{not important} near the particle-hole symmetry point, at
least for an accurate prediction of the resonance position.
However, this is not clear from the start and required a careful numerical \New{examination}.
Furthermore, our
kinetic equations (\ref{eq:spr-kineqdiscuss}) are compactly \New{expressed in equations for physically meaningful observable averages.}\\
\New{\emph{Other methods.}}
Several other works dealing with noncollinearly polarized leads
employ completely different techniques, such as a Green's function approach in
the noninteracting approximation \cite{Pedersen05}, in a Hartree-Fock
approximation \cite{Rudzinski05,Sergueev02,Wilczynski05}, or restricted to
zero bias \cite{Fransson05}. As these works do \New{not employ} kinetic
equations, a direct comparison of the results is more difficult. Some of these
studies address different exchange field effects also for noncollinear
polarizations; \cite{Fransson05,Rudzinski05} yet, a sharp resonance has not
been reported there.\\
\New{
Thus, even though we investigate in this paper the well-studied
Anderson QD model with noncollinear ferromagnets, our technically \new{advanced} analysis
gives us access to a parameter regime for which
reliable predictions were hardly possible before.
This allows us to go beyond previous works.
The reason that our spin resonance without spin splitting has been
overlooked so far is that it requires the careful treatment of the combination of
(i) slow decoherence of the spin in the Coulomb blockade regime, (ii) the
degeneracy of the spin levels allowing the coherent evolution to be dominated
by the exchange field, and (iii) complete rotational symmetry breaking by
noncollinearly polarized ferromagnet.}

\subsection{Solving the quantum master equation in the Coulomb blockade
regime}\label{app:crossover}
As explained in Sec. \ref{sec:notation}, in the Coulomb
blockade regime the next-to-leading order $\Gamma^2$ contributions can
dominate over the leading-order $\Gamma$ contributions. This also requires
careful consideration when solving Eq. (\ref{eq:spr-kineqdiscuss}) for the
occupation probabilities $p_n$ and the average spin $\vec{S}$: \new{To solve}
the kinetic equations one could perform a systematic perturbation expansion
not only for the kernels but also for the probabilities $p_n = p_n^{(0)} +
p_n^{(1)} +$... and the spin $\vec{S} = \vec{S}^{(0)} + \vec{S}^{(1)} +
\ldots$ in orders of $\Gamma$, and solve Eq. (\ref{eq:kineq}) then order by
order in $\Gamma$. This has the advantage that the current is evaluated
consistently to a given order in $\Gamma$. This procedure works well as long
as lowest-order $\Gamma$ tunneling processes (sequential tunneling) are
present but fails in the Coulomb blockade regime where sequential tunneling is
exponentially suppressed and cotunneling dominates. This is particularly
important for the results obtained for infinite $U$, shown in
Fig. \ref{fig:timedep}. Therefore, we use an alternative procedure in which
only the kernels (but not the probabilities and the spin) are expanded in
powers of $\Gamma$. It is, thus, the {\tmem{kernels}} that are to be
consistently evaluated to a given order in $\Gamma$: not the density operator
or observables such as the current. This issue has been thoroughly discussed
\New{elsewhere} for our model  \cite{Weymann05} \New{but also in a more general context \cite{Leijnse08a} including,
e.g., vibrational degrees of freedom on the QD.} Although
the current we obtain may comprise terms of order $\Gamma^3$, we checked that
the spin resonance is clearly not an artifact of those terms. By
varying $\Gamma$, the resonance current is found to scale \new{maximally
as $\Gamma^2$ or a lower power but not as $\Gamma^3$.}
\subsection{Incorporation of superconducting terminal}\label{app:kineqsc}
We comment here on the results that we show in Fig. \ref{fig:resonance_sc} when adding a proximal superconductor to the setup. To
simplify the \New{analytical calculations (which are already quite
involved without a superconductor)}, we included here only the
leading-order $\Gamma$
contribution in the tunneling rates. Consistent with this, the charging energy
$U$ has been chosen of moderate size there. There are several reasons why this
simplification does not affect the conclusions we draw from
Fig. \ref{fig:resonance_sc} that concern only the resonance
{\tmem{position}}. First, we note that the effect of the superconductor is
clearly visible when moving into the Coulomb diamond but still within the
thermal broadening window of $4 T$ around the single-electron tunneling resonances in
Fig. \ref{fig:resonance_sc}. Here, a leading-order $\Gamma$
calculation gives reliable predictions without any question. Second, we note
that this regime covers quite a large part of Fig. \ref{fig:resonance_sc} since the presence of the superconductor reduces the
size of the effective Coulomb diamond in the stability diagram, as one sees
from comparing Figs. \ref{fig:resonance} and
\ref{fig:resonance_sc}. The exponential suppression of the
$O (\Gamma)$ rates is thus attenuated, but it may still be strong near the
particle-hole symmetry point. Here, one should in principle include $O
(\Gamma^2)$ corrections. However -- and this is our third point -- by
comparing results of $O (\Gamma)$ [not shown here], and O($\Gamma^2$)
[Fig. \ref{fig:resonance}] for the same parameters without
superconductor, we know that the resonance is not diminished, as clearly
demonstrated by Fig. \ref{fig:resonance}, but only slightly
broadened due to the additional spin decay introduced by cotunneling [c.f.
Sec. \ref{sec:shape}]. Once the resonance appears, its
position is determined by the {\tmem{first}}-order exchange field [cf.
Sec. \ref{sec:overview}], modified by the proximal
superconductor [cf. Sec. \ref{sec:sc}], the effect we
wished to illustrate here. The cotunneling corrections are not needed to draw
a conclusion about the resonance {\tmem{position}}.
\subsection{Non-Markovian corrections}\label{app:nonmarkov}
Finally, we comment on the validity of the Markovian approximation underlying
our kinetic equations (\ref{eq:spr-kineqdiscuss}) for our study of the
time-dependent pulsing scheme. 
 To study time-dependent problems in
the Coulomb blockade regime, one must in principle also include non-Markovian
corrections into the kernel \cite{Splettstoesser10}. However, non-Markovian
corrections appear only as modifications of the {\tmem{next-to-leading order}}
contributions. Thus, non-Markovian corrections do not affect the exchange
field, which is dominated by leading-order terms and determines the position
of the spin resonance and the frequency of the spin precession. On the
contrary, the corrections do alter the spin-decay tensor $\mathcal{R}_S$ and
thereby the time constant of the damped spin oscillations. In spite of this,
the latter will still be of O($\Gamma^2 / U$) in the Coulomb blockade regime,
which we have identified as the crucial requirement for the underdamped spin
precession. Including non-Markovian corrections is, hence, required only for a
quantitative analysis but not to demonstrate the viability of an underdamped
spin precession, which is our aim here. It should be noted that if such
accuracy is desirable, other spin-decay mechanisms should also be taken into
account (see Sec. \ref{sec:estimate}), which is clearly
beyond the scope of this work.

\section{Particle-hole symmetry}\label{sec:spr-phs}
\New{In Sec. \ref{sec:position}, our discussion of the resonance position applied} only to the left half
of the Coulomb diamond, i.e., for gate voltages $\delta = U + 2 \varepsilon> 0$; cf.
Fig. \ref{fig:resonance}. \New{Here, we} show that the resonance
extends point-symmetrically with respect to the particle-hole symmetry point
$(\delta, V_b) = (0, 0)$. In the region \New{to} the right of this point, where
$\delta < 0$, the resonance condition requires the exchange field to be
perpendicular to the {\tmem{drain}} polarization:
\begin{eqnarray}
  & \begin{array}{lll}
    \vec{B} \cdot \hat{\vec{n}}_d & = & 0
  \end{array} & (\delta < 0) .  \label{eq:resonancedrain}
\end{eqnarray}
This condition is fulfilled for {\tmem{negative}} resonant bias $V^{\ast}_b <
0$. Thus, the drain refers here to the same physical electrode as the source
in Eq. (\ref{eq:resonance}) because changing the sign of the bias exchanges
the role of source and drain.\\ 
Equation (\ref{eq:resonancedrain}) can be
understood physically as follows: For $\delta < 0$, the electrochemical
potential of the leads is closer to that for the doubly occupied QD and
therefore the current predominantly involves the doubly occupied QD state.
Consequently, when an electron leaves the QD, it leaves behind a {\tmem{hole}}
polarized along $\hat{\vec{n}}_d$. However, an accumulation of hole spins can be
efficiently prevented by the exchange field $\vec{B}$ if the latter is
directed perpendicular to $\hat{\vec{n}}_d$, that is, if condition
(\ref{eq:resonancedrain}) is fulfilled.\\
In analogy to Eq. (\ref{eq:adivq}), condition (\ref{eq:resonancedrain}) can be
recast as
\begin{eqnarray}
  \frac{a'}{q'} & = & 1 \label{eq:adivqdrain},
\end{eqnarray}
with
\begin{eqnarray}
  a' & = & \frac{\Gamma_s n_s \cos (\alpha)}{\Gamma_d n_d },
\end{eqnarray}
and
\begin{eqnarray}
  q' & : = & \frac{\phi_d (\varepsilon) - \phi_d (\varepsilon + U)}{\phi_s
  (\varepsilon) - \phi_s (\varepsilon + U)} \text{ \ } (\delta < 0), 
\end{eqnarray}
where the spin injection asymmetry ratio $a'$ is defined differently as in Eq.
(\ref{eq:adivq}). Equation (\ref{eq:adivqdrain}) complies with Eq.
(\ref{eq:adivq}): Mapping $(\delta, V_b) \rightarrow (- \delta, - V_b)_{}$, we
have to replace $q \rightarrow 1 \text{/} q'$ and $a \rightarrow 1 \text{/}
a'$ since the roles of source and drain are interchanged. It is therefore
sufficient to discuss only the case $\delta > 0$ \New{as we did in Sec. \ref{sec:position}} as all results hold for
$\delta < 0$ accordingly by reversing the signs of $\delta$ and $V_b$.\\

\end{appendix}

\bibliographystyle{apsrev}
\end{document}